\documentclass[final,1p,times]{elsarticle}

\usepackage{graphicx}
\usepackage{slashbox}

\usepackage{amssymb}

\usepackage[nodots,nocompress]{numcompress}

\usepackage[]{subfigure}
\biboptions{comma,square}

\journal{Astroparticle Physics}
\begin{document}

\begin{frontmatter}



\title{The analog Resistive Plate Chamber detector \\of the ARGO-YBJ experiment}

\author[1,2]{B.~Bartoli}
\author[3,4]{ P.~Bernardini}
\author[5]{ X.J.~Bi}
\author[8]{ P.~Branchini}
\author[8]{ A.~Budano}
\author[10,11]{ P.~Camarri}
\author[5]{ Z.~Cao}
\author[11]{ R.~Cardarelli}
\author[1,2]{ S.~Catalanotti}
\author[5]{S.Z.~Chen}
\author[12]{ T.L.~Chen}
\author[4]{ P.~Creti}
\author[13]{ S.W.~Cui}
\author[14]{ B.Z.~Dai}
\author[3,4]{ A.~D'Amone}
\author[12]{ Danzengluobu}
\author[3,4]{ I.~De Mitri}
\author[1,2]{ B.~D'Ettorre Piazzoli}
\author[1,2]{ T.~Di Girolamo}
\author[11]{ G.~Di Sciascio}
\author[15]{ C.F.~Feng}
\author[5]{ Zhaoyang Feng}
\author[16]{ Zhenyong Feng}
\author[5]{ Q.B.~Gou}
\author[5]{ Y.Q.~Guo}
\author[5]{ H.H.~He}
\author[12]{ Haibing Hu}
\author[5]{ Hongbo Hu}
\author[1,2]{ M.~Iacovacci\corref{cor1}}
\cortext[cor1]{corresponding author, email: iacovacci@na.infn.it}
\author[10,11]{ R.~Iuppa}
\author[16]{ H.Y.~Jia}
\author[12]{ Labaciren}
\author[12]{ H.J.~Li}
\author[6,7]{ G.~Liguori}
\author[5]{ C.~Liu}
\author[14]{ J.~Liu}
\author[12]{ M.Y.~Liu}
\author[5]{ H.~Lu}
\author[5]{ L.L.~Ma}
\author[5]{ X.H.~Ma}
\author[3,4]{ G.~Mancarella}
\author[8,17]{ S.M.~Mari}
\author[3,4]{ G.~Marsella}
\author[3,4]{ D.~Martello}
\author[2]{ S.~Mastroianni}
\author[8,17]{ P.~Montini}
\author[12]{ C.C.~Ning}
\author[3,4]{ M.~Panareo}
\author[3,4]{ L.~Perrone}
\author[8,17]{ P.~Pistilli}
\author[8]{ F.~Ruggieri}
\author[7]{ P.~Salvini}
\author[10,11]{ R.~Santonico}
\author[5]{ P.R.~Shen}
\author[5]{ X.D.~Sheng}
\author[5]{ F.~Shi}
\author[4]{ A.~Surdo}
\author[5]{ Y.H.~Tan}
\author[18,19]{ P.~Vallania}
\author[18,19]{ S.~Vernetto}
\author[19,20]{ C.~Vigorito}
\author[5]{ H.~Wang}
\author[5]{ C.Y.~Wu}
\author[5]{ H.R.~Wu}
\author[15]{ L.~Xue}
\author[14]{ Q.Y.~Yang}
\author[14]{ X.C.~Yang}
\author[5]{ Z.G.~Yao}
\author[12]{ A.F.~Yuan}
\author[5]{ M.~Zha}
\author[5]{ H.M.~Zhang}
\author[14]{ L.~Zhang}
\author[15]{ X.Y.~Zhang}
\author[5]{ Y.~Zhang}
\author[5]{ J.~Zhao}
\author[12]{ Zhaxiciren}
\author[12]{ Zhaxisangzhu}
\author[16]{ X.X.~Zhou}
\author[16]{ F.R.~Zhu}
\author[5]{ Q.Q.~Zhu}
\author[9]{and G.~Zizzi}\par
\author[] {\\(The ARGO-YBJ Collaboration)}


 \address[1]{Dipartimento di Fisica dell'Universit\`a di Napoli
                  ``Federico II'', Complesso Universitario di Monte
                  Sant'Angelo, via Cinthia, 80126 Napoli, Italy.}
 \address[2]{Istituto Nazionale di Fisica Nucleare, Sezione di
                  Napoli, Complesso Universitario di Monte
                  Sant'Angelo, via Cinthia, 80126 Napoli, Italy.}
 \address[3]{Dipartimento Matematica e Fisica "Ennio De Giorgi",
                  Universit\`a del Salento,
                  via per Arnesano, 73100 Lecce, Italy.}
 \address[4]{Istituto Nazionale di Fisica Nucleare, Sezione di
                  Lecce, via per Arnesano, 73100 Lecce, Italy.}
 \address[5]{Key Laboratory of Particle Astrophysics, Institute
                  of High Energy Physics, Chinese Academy of Sciences,
                  P.O. Box 918, 100049 Beijing, P.R. China.}
 \address[6]{Dipartimento di Fisica dell'Universit\`a di
                  Pavia, via Bassi 6, 27100 Pavia, Italy.}
 \address[7]{Istituto Nazionale di Fisica Nucleare, Sezione di Pavia,
                  via Bassi 6, 27100 Pavia, Italy.}
 \address[8]{Istituto Nazionale di Fisica Nucleare, Sezione di
                  Roma Tre, via della Vasca Navale 84, 00146 Roma, Italy.}
 \address[9]{Istituto Nazionale di Fisica Nucleare - CNAF, Viale
                  Berti-Pichat 6/2, 40127 Bologna, Italy.}
 \address[10]{Dipartimento di Fisica dell'Universit\`a di Roma ``Tor Vergata'',
                   via della Ricerca Scientifica 1, 00133 Roma, Italy.}
 \address[11]{Istituto Nazionale di Fisica Nucleare, Sezione di
                   Roma Tor Vergata, via della Ricerca Scientifica 1,
                   00133 Roma, Italy.}
 \address[12]{Tibet University, 850000 Lhasa, Xizang, P.R. China.}
 \address[13]{Hebei Normal University, Shijiazhuang 050016,
                   Hebei, P.R. China.}
 \address[14]{Yunnan University, 2 North Cuihu Rd., 650091 Kunming,
                   Yunnan, P.R. China.}
 \address[15]{Shandong University, 250100 Jinan, Shandong, P.R. China.}
 \address[16]{Southwest Jiaotong University, 610031 Chengdu,
                   Sichuan, P.R. China.}
 \address[17]{Dipartimento di Fisica dell'Universit\`a ``Roma Tre'',
                   via della Vasca Navale 84, 00146 Roma, Italy.}
 \address[18]{Osservatorio Astrofisico di Torino dell'Istituto Nazionale
                   di Astrofisica, via P. Giuria 1, 10125 Torino, Italy.}
 \address[19]{Istituto Nazionale di Fisica Nucleare,
                   Sezione di Torino, via P. Giuria 1, 10125 Torino, Italy.}
 \address[20]{Dipartimento di Fisica dell'Universit\`a di
                   Torino, via P. Giuria 1, 10125 Torino, Italy.}





\begin{abstract}
The ARGO-YBJ experiment has been in stable data taking from November 2007 till February 2013 at the YangBaJing Cosmic Ray Observatory (4300 m a.s.l.). The detector consists of a single layer of Resistive Plate Chambers (RPCs) (6700 m\textbf{$^2$}) operated in streamer mode. The signal pick-up is obtained by means of strips facing one side of the gas volume. The digital readout of the signals, while allows a high space-time resolution in the shower front reconstruction, limits the measurable energy to a few hundred TeV.
In order to fully investigate the 1-10 PeV region, an analog readout has been implemented by instrumenting each RPC with two large size electrodes facing the other side of the gas volume. Since December 2009 the RPC charge readout has been in operation on the entire central carpet ($\sim$ 5800 m$^2$).
In this configuration the detector is able to measure the particle density at the core position where it ranges from tens to many thousands of particles per m$^2$.
Thus ARGO-YBJ provides a highly detailed image of the charge component at the core of air showers.
In this paper we describe the analog readout of RPCs in ARGO-YBJ and discuss both the performance of the system and the physical impact on the EAS measurements.
\end{abstract}


\end{frontmatter}








\section{Introduction}

The ARGO-YBJ experiment, promoted and funded by the Italian Institute for Nuclear Physics (INFN) and by the Chinese Academy of Sciences (CAS) in the framework of the Italy–China scientific cooperation, has operated at the YangBaJing Cosmic Ray Laboratory (Tibet, PR China, 4300 m a.s.l., 606 g/cm$^2$) from November 2007 till February 2013.
As typical of an air shower array apparatus, it has benefited from a large field of view ($\sim$ 2 sr) and a high duty cycle ($>85\%$), that allowed a continuous monitoring of the sky in the declination band from -10$^\circ$ to 70$^\circ$. The two key features, namely the dense sampling active area and the operation at high altitude, allowed the study of the cosmic radiation at an energy threshold of a few hundred GeV.

ARGO-YBJ has been proposed as an experiment capable of investigating a wide range of fundamental issues in Cosmic Ray and Astroparticle Physics at a relatively low energy threshold:  high energy $\gamma$-ray astronomy, at an energy threshold of a few hundred GeV;  search for emission of $\gamma$-ray bursts in the full GeV-TeV energy range;  Cosmic Ray (CR) physics (energy spectrum, chemical composition, $\bar{p}/p$ ratio measurement, shower space-time structure, multicore events, p-air and pp cross section measurement) starting from TeV energies;  Sun and heliosphere physics above  1 GeV.

The detector could measure in different energy ranges, according to the way of operation, which essentially means the minimum measurable spatial density of particles. In scaler mode, in order to lower the energy threshold down to 1 GeV \cite{Aielli(2008)}, the total counts on sub-units of 44 m$^2$ are measured every 0.5 s, with limited information on both space distribution and arrival direction of the detected particles. In shower mode, position and arrival time of the charged particles are measured, then the arrival direction of the primary particle is determined by a fast timing technique \cite{Aielli 2009b}.

In shower mode ARGO-YBJ can determine the number of particles in the shower either by counting the number of fired pick-up strips (digital readout) or measuring the total charge induced by the particles passing through the detector (analog readout). The maximum particle density which can be measured is different for the two cases, simply depending on geometry in the first case (23 strips/m$^2$) or on electronics in the second case (up to many 10$^4$ /m$^2$), corresponding to a maximum detectable energy of a few hundred TeV in the first case and to many PeV in the second case.

From 2007 to the end of operation ARGO-YBJ has been operated in scaler mode and in shower mode. At December 2009 the analog readout was made operational  so increasing the maximum achievable energy up to many PeV. Due to the dimension of the detector, the gamma astronomy studies are limited to energies $ < $ 20 TeV \cite{ARGO Crab}, well addressed by the digital readout (see \cite{sky survey} and reference therein). Therefore, in what follows, we will consider the major results obtained so far by  ARGO-YBJ on the CR physics and the new opportunities offered by the analog mode of detector operation.

\begin{itemize}
 \item
Using the strip multiplicity spectrum measured by ARGO-YBJ the light-component (proton and helium) spectrum of the primary CR in the energy region 5-200 TeV has been evaluated by using a Bayesian approach \cite{PRD_light}, so bridging for the first time direct and indirect measurements below 100 TeV, region not accessible by other EAS experiments.
The ARGO-YBJ data agree remarkably well (within about 15$\%$) with the values obtained by adding up the proton and helium fluxes measured by CREAM \cite{{CREAM2009},{CREAM2011}}, both concerning the total intensities and the spectrum slope, so confirming the CREAM finding that the proton and helium spectra, from 2.5 to 250 TeV, are both flatter compared with the lower energy measurements.
Later on the energy spectrum of proton and helium has been measured below the so-called
"knee", from 100 to 700 TeV,  by using a hybrid experiment\cite{CPC 2014} made of a wide field-of-view Cherenkov telescope \cite{WFCTA} and the ARGO-YBJ carpet.
The light component has been well separated from other CR components by using a multi-parameter technique
where the analog information has been used for the first time. A highly uniform energy resolution of about 25$\%$ was achieved throughout the whole energy range (100-700 TeV). The measured spectrum agrees in both spectral index and absolute flux with the spectrum obtained by ARGO-YBJ alone in the lower energy range from 5 TeV to 200 TeV.
Since proton and helium nuclei are the bulk of the CR at energies below the knee (3$\times$ 10$^{15}$ eV), the study of their spectrum in this energy region is of primary importance.

\item
  Basing on the CR flux attenuation for different atmospheric depths, i.e. zenith angles, and exploiting the high detector accuracy in reconstructing the shower properties at ground, ARGO-YBJ  has measured the production cross section between 1 - 100 TeV CR protons and Air nuclei \cite{(pp-xsec 2009)}.
The analysis results have been also used to estimate the total proton-proton cross section at center-of-mass energies between 70 and 500 GeV. The total proton-proton cross section has then been inferred from the measured proton-Air production cross section by using the Glauber theory.
The result is consistent with the general trend of experimental data, favoring an asymptotic ln$^2$(s) rise of the cross section.

 \item
ARGO-YBJ has reported \cite{MSA PRD} the observation of anisotropic structures on a medium angular scale as wide as $\sim$ 10$^\circ$ - 45$^\circ$, in the energy range 10$^{12}$ - 10$^{13}$ eV. The intensity spans from 10$^{-4}$ to 10$^{-3}$, depending on the selected energy interval and sky region. For the first time, the observation of new MSA structures throughout the right ascension region 195$^\circ$ - 290$^\circ$ has been observed with a statistical significance above 5 s.d.. At higher energies both the EAS-TOP \cite{Aglietta2009} and IceCube \cite{Abbasi2011} experiments observed significant anisotropy around 400 TeV. At this energy, the signal looks quite different from the modulation observed up to  50 TeV, both in amplitude and phase; it suggests that the global anisotropy may be the superposition of different contributions from phenomena at different distances from the Earth.

 \item

In the ARGO-YBJ experiment, for the first time, the geomagnetic field (GeoMF) lateral stretching is observed \cite{GeoMF PRD} at small distances from the shower core. This effect has been suggest by Cocconi \cite{Cocconi1954} who supposed that the lateral displacement induced by the Earth’s magnetic field is not negligible with respect to the Coulomb scattering when the shower is young.
The non-uniformity in the azimuthal distribution  has been deeply studied by ARGO-YBJ concluding that it is well described by two harmonics: the first one of about 1.5$\%$ and the second one of about $0.5\%$. The first harmonic is due to the GeoMF; the second one is the sum of the magnetic and detector effects.
The phase of the first harmonic can be used as a marker of the absolute pointing of an EAS array and, if any, can be used to apply a simple correction to the absolute pointing. The ARGO-YBJ experimental results represent the first evidence of the charge density reduction near the EAS axis due to the GeoMF.

\end{itemize}

With the analog readout in operation on 5800 m$^2$ ARGO-YBJ can extend its CR physics results at higher energies ($\rm{>100}$ TeV), so reaching the knee region of the CR spectrum, and cope with new  physics items.
The energy spectrum of all CR primaries, the all-particle spectrum, will be measured up to the knee  region and the evolution of the different components will be investigated too with special attention to the light component. In this respect we notice the methodological difference with respect to the sampling apparatus, in fact ARGO-YBJ will use only information provided by the very near region of the core, that essentially means truncated size, local age and particle density at the core position.

Improvements are expected on the p-Air cross section measurement, that is the extension to center-of-mass energies in the TeV region. The possibility of performing this measurement actually strongly relies on the knowledge of the chemical composition around the knee, specifically of the light component, which is still matter of debate and investigations. So the light component and the p-Air cross section measurements look as intimately correlated and will represent a key item in the ARGO-YBJ investigations. Considering the geometry of the detector, the core region is measured with unprecedented detail and the very forward region of the interaction can be studied. It will be interesting to see how well the different models will describe the data at higher energies. In this context it is worthwhile to note that the proton-air inelastic cross section measured by the ARGO-YBJ experiment in the energy range 1-100 TeV has been found in good agreement with the values set in the CORSIKA/QGSJET code, and that, according to the results shown in\cite{PRD_light}, the QGSJET and SYBILL models provide the same description of the longitudinal development of the shower below a few hundred TeV.

Due to the size of the detector, then to the statistics of collected data, it will be almost impossible to detect the CR anisotropy signal at high energies especially considering the level of 10$^{-4}$ reported \cite{Aglietta2009}, even using not perfectly internal events.

The study of the GeoMF effect on the shower development are expected to be performed at higher energies, in fact an azimuthal modulation was already observed not only at the Yakutsk array for EAS with energies above 50 PeV \cite{Ivanov1999} and at the ALBORZ Observatory for energies above 100 TeV \cite{Bahmanabadi2002},  but also in radio experiments \cite{Ardouin2009}.

Shower events with multiple core structure, or multicore events, have been observed and investigated by many experiments with different techniques, namely emulsion chambers \cite{MEC} and hadronic calorimeter \cite{EasTop}. Those events are typically explained within the framework of jet production, which is essentially provided by the leading particle interaction with the Air target nuclei, the separation between the cores being related to the $p_T$ of the leading particle. The ARGO-YBJ carpet with its huge surface ($\sim 5800 m^2$) allows the measurement of events with cores separated up to 100 m, so providing a good opportunity to measure very high $p_T$ events \cite{Zhao}. Events at high $p_T$ are a perfect tool to investigate hadronic interaction models or even new physics. In this respect also the study of double front showers, at high energy, would be a check for some exotic physics \cite{Calabrese}.
Measuring the shower core region will allow the study of local clustering phenomena of  particles in the shower that could have a strong correlation with the mass of the primary hadron; in fact this phenomena should be enhanced around the core region \cite{MinZha}.
In this paper the detector layout is described in $\S$2 while the main features of the analog signal from the RPC are discussed in $\S$3.  The $\S$4 copes with the question of the intrinsic linearity of the RPC detector and results are reported on a specific test-beam. Then in $\S$5 the trigger of the experiment and the related electronics are introduced. The architecture of the charge readout system and the local trigger that enables the charge readout on the single Cluster are discussed in $\S$6, while in $\S$7 the logic operations of the charge readout system are presented focusing on synchronization, data collection and data packing. The main operations of the DAQ control system concerning the analog readout (managing, calibration and monitoring) are described in $\S$8. Then, in $\S$9, the calibration chain of the analog information is presented and the stability of the detector is discussed. Checks of consistency and the final performance of the analog readout of the detector are shown in $\S$10.
A summary of the experimental results is given in $\S$11 while the conclusions are drawn in $\S$12.

\section{The ARGO-YBJ detector}
\begin{figure*}[th!]
\centering
\includegraphics[width=5.0in]{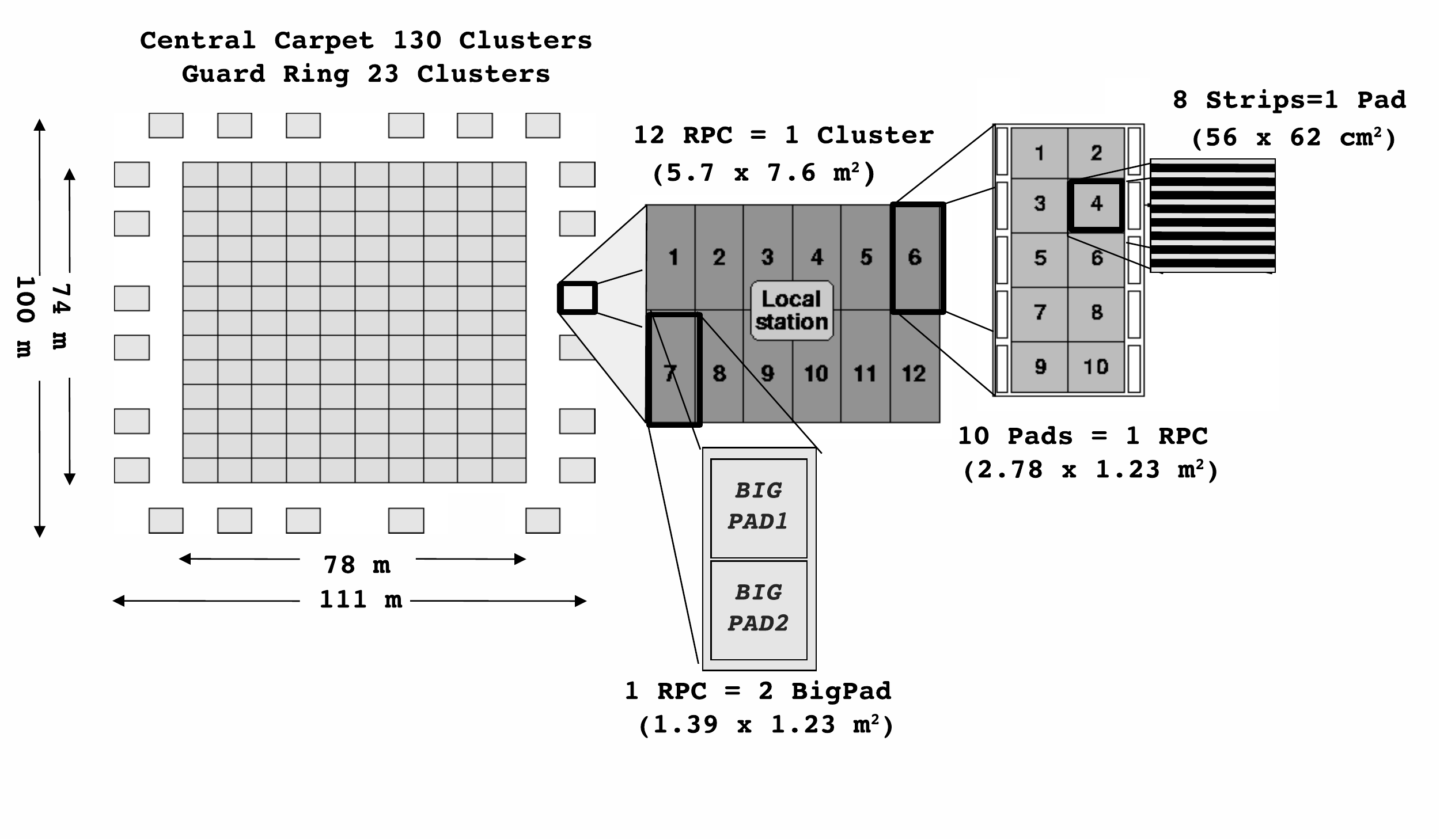}
\caption{The ARGO-YBJ detector setup. The Cluster (12 RPCs) is the basic module. The Local Station (LS) is the DAQ unit providing local trigger and readout.}
\label{lay-out-ARGO}
\end{figure*}
The ARGO-YBJ detector \cite{aielli06}, hosted in a building at the YangBaJing Cosmic Ray Observatory, consists of a central carpet $\sim $ 78$\times $74 m$^{2}$, made of a single layer of Resistive Plate Chambers (RPCs) with $\sim $93$\%$ active coverage, enclosed by a guard ring partially instrumented ($\sim $20$\%$)  up to $\sim $110$\times $100 m$^{2}$. The apparatus has a modular structure, the basic data acquisition unit being a Cluster (7.6$\times $ 5.7 m$^{2}$), made of 12 RPCs (1.25 $\times $2.80 m$^{2}$ each). The full detector has 153 Clusters (130 in the central carpet, 23 in the guard ring) with a total active surface of about 6700 m$^{2}$, as shown in Fig\ref{lay-out-ARGO}. Each RPC is read out by means of 80 pick-up strips (61.8$\times $6.75 cm$^{2}$, the spatial pixels) facing one side of the gas volume. The fast-OR signal of 8 contiguous strips defines the logical pad (61.8 $\times $ 55.6 cm $^{2}$, the time pixel) which is used for timing and triggering purposes. Any manifold coincidence of fired pads of the central carpet ($N_{pad}$) above a given multiplicity threshold ($N_{trig}$), namely $N_{pad} \geq N_{trig}$  in a time window of 420 ns, implements the inclusive trigger that starts the event data acquisition. The apparatus, in its full configuration of 153 Clusters, has been in smooth and stable data taking since November 2007 till February 2013 with a duty cycle $\geq 86 \%$.
The trigger threshold is $N_{pad}\geq 20 $ ($\rm{E_0} \geq 300 (900)$ GeV for primary photons (protons)) and the corresponding trigger rate is $\sim $ 3.5 kHz.
The high granularity of the detector and its time resolution provide a detailed three-dimensional reconstruction of the shower front. The digital pick-up of the RPC, which has a density of 23 strips/m$^{2}$, can be used to study the primary spectrum up to energies of a few hundred TeV; above these energies its response saturates \cite{saggese}. In order to extend the measurable energy range and fully investigate PeV energies, where particle densities at the core position are larger than 10$^{3}$/m$^{2}$, each RPC has also been equipped with two large size electrodes of dimensions 1.23 $\times $ 1.39 m$^{2}$. These pick-up electrodes, called Big Pads (BP), face the other side of the RPC gas volume (see Fig.\ref{RPCLayout}) and provide a signal whose amplitude is expected to be proportional to the number of charged particles impinging on the detector. \\
The BP and the 40 corresponding strips see the same signals provided by particles impinging on the RPC and detected by it.
The RPCs are operated in streamer mode, with a gas mixture of Argon (15$\%$), Isobutane (10$\%$) and Tetraflouroethane (75$\%$). The operating voltage is 7.2 kV. This setting provides a typical efficiency $>$ 95$\%$ with an intrinsic time resolution of about 1 ns and a minimum ionizing particle (m.i.p.) signal on the BP of $\sim$2 mV.\\

\section{Chamber Layout and Signal Characteristics}
The chamber layout and the results concerning the digital read-out have already been discussed in \cite{aielli06}.
\begin{figure}[th!]
  \begin{center}
    \includegraphics[width=4.0in]{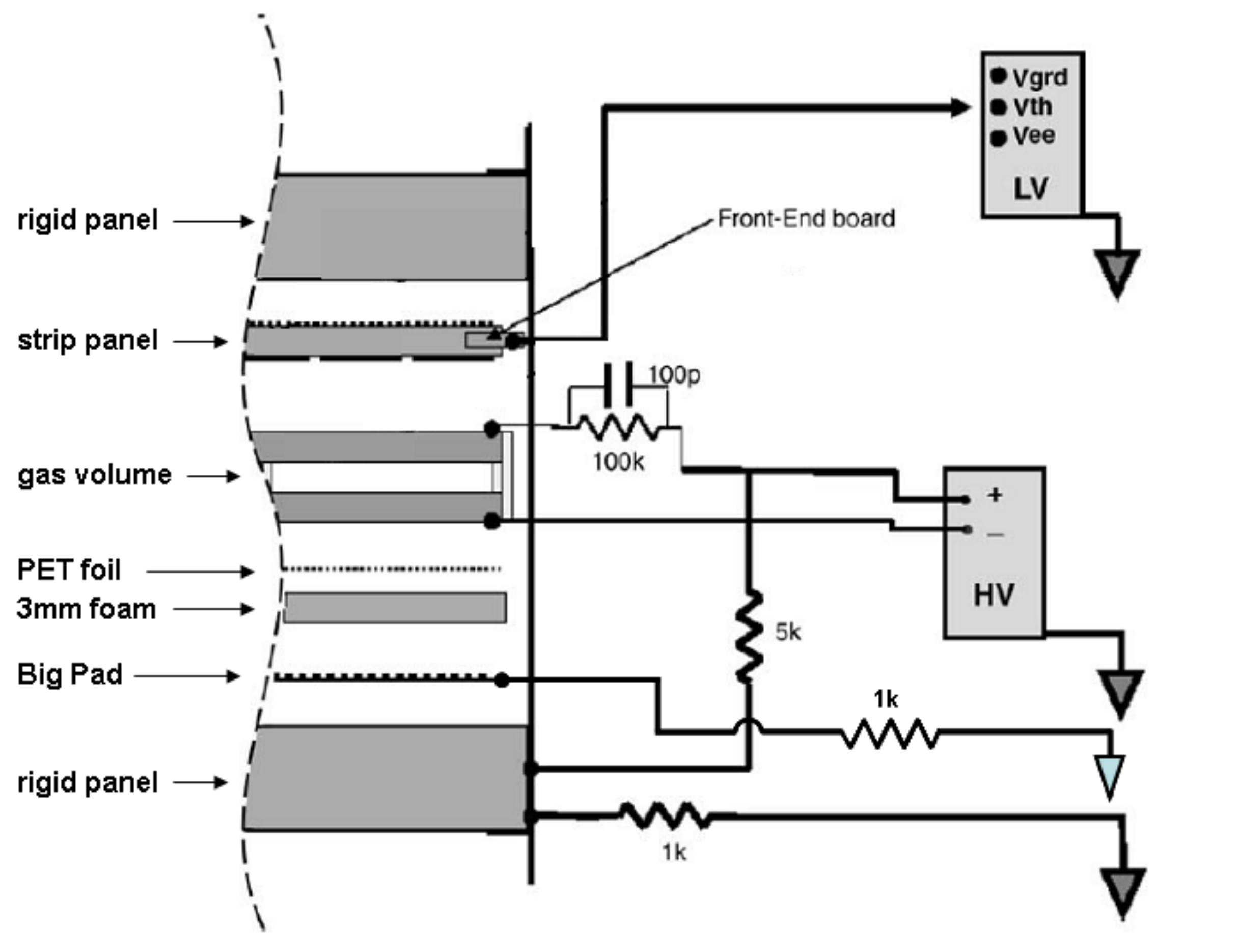}
  \end{center}
    \caption{Layout of the Argo-YBJ RPC with  High Voltage, Low Voltage and grounding connections (see \cite{aielli06} for details).}
    \label{RPCLayout}
\end{figure}
According to the layout shown in Fig.\ref{RPCLayout} the BP is a large pick up electrode grounded through a 1 k$\Omega$ resistor in order to avoid slow charging, typical of a floating electrode, followed by a breakdown discharge with possible electronics damage.\\
It is made by a copper foil (17 $\mu$m thick) glued on a PET foil (190 $\mu$m).

\begin{figure}[th!]
  \begin{center}
\includegraphics[width=4.0in]{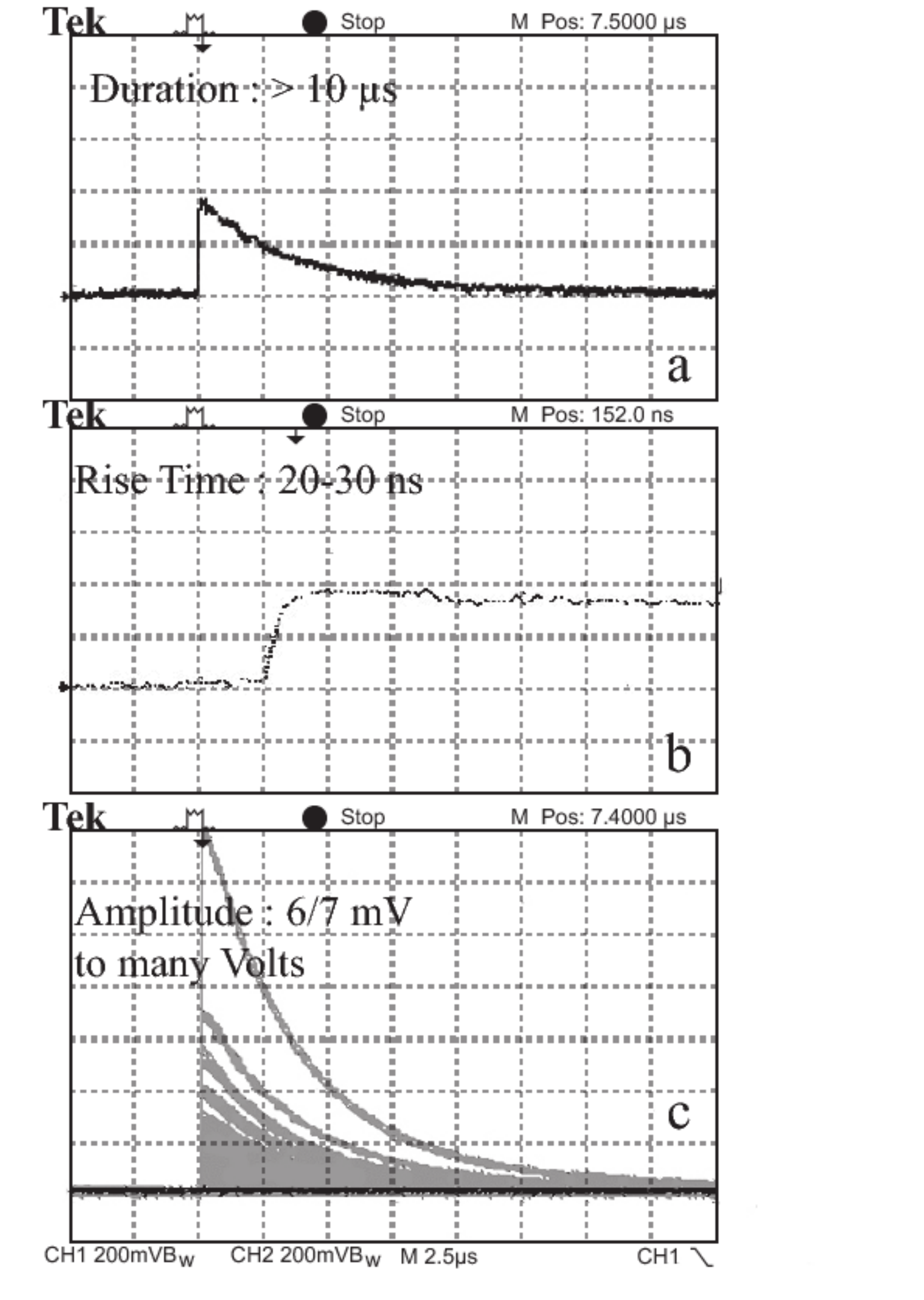}
  \end{center}
    \caption{Typical signal from the Big Pad  on a $50 ~ \Omega$ load. The signal is reported in a) with a horizontal scale of $2.5~ \mu s$ and in b) with a horizontal scale of $100~ ns$. In both cases the vertical scale is $10 ~mV$. In Fig.\ref{TypicalSignal}c the vertical scale is $200 ~mV$ (horizontal scale $2.5~ \mu s$) with the oscilloscope set at infinite persistence and triggered for $20$ minutes by particles crossing the chamber.}
    \label{TypicalSignal}
\end{figure}

A typical signal of the BP is shown in Fig.\ref{TypicalSignal}: it has a rise time in the range of a few tens of ns and a discharge time of about 5 $\mu$µs; the amplitude of the signal, on a $50~ \Omega$ load, ranges from milliVolts to tens of Volt.
\begin{figure}[t!]
  \begin{center}
    \includegraphics[width=4.0in]{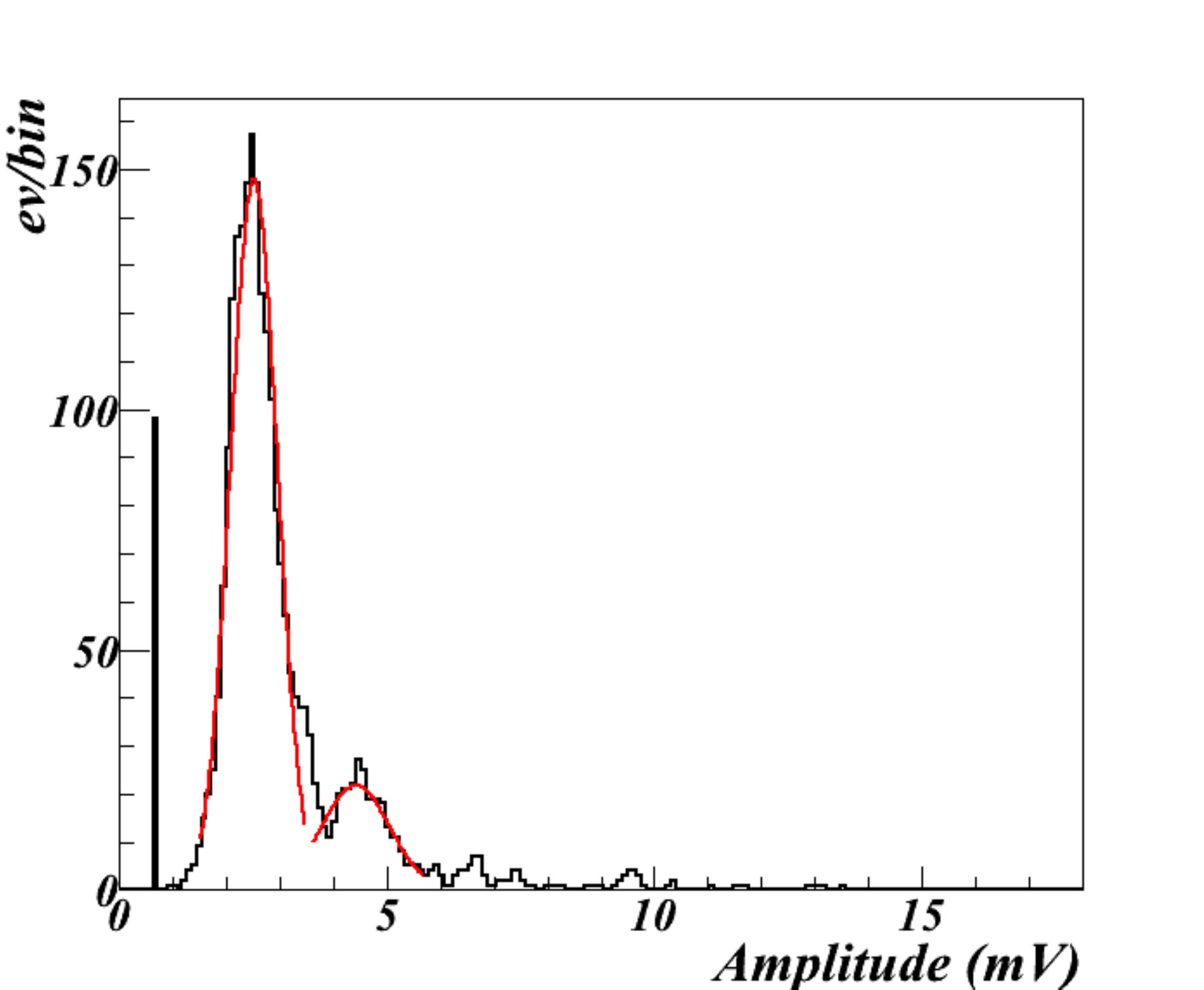}
  \end{center}
  \vspace{-0.3pc}
   \caption{Amplitude distribution of the Big Pad signals induced by m.i.p.,  at 7.2 kV, as measured at the experimental site of YangBaJing . The pedestal (the BP electronic noise) as well as the single and double streamer peaks are clearly shown. The bin size is $\sim$0.1 mV.The two curves represent the best gaussian fits to the first and second peak of the distribution.}
   \label{Amplitude distribution}
\end{figure}
In Fig.\ref{Amplitude distribution} the amplitude distribution of the signals generated by a m.i.p., as measured at the experiment site, is reported along with the gaussian fits to the first and second peak of the distribution. The "single particle" trigger was provided by a small telescope \cite{santonico00} made of three RPCs of 50$\times $50 cm$^{2}$, two above and one below the chamber under measurement, with about 1 cm of lead put on top of the lower RPC, and the triple coincidence of their signals used to select a m.i.p.. The amplitude distribution of the signal provided by the BP on a $50~ \Omega$ load, measured by a ChargeMeter board specifically designed to this aim,  shows a first peak at $\sim 2.5~mV$ which represents about 90$\%$  of the sample and corresponds to one streamer, with $\sigma \sim 0.4~mV$, and a second peak at about $4.4~mV$ with $\sigma \sim 0.6~mV$) , which corresponds to two streamers. The  first populated bin shows a spike at 0.7 mV which corresponds to electronic noise, in fact its content is consistent with the overall RPC inefficiency, $\sim$ 3$\%$.
\begin{figure}[t!]
  \begin{center}
    \includegraphics[width=4.0in]{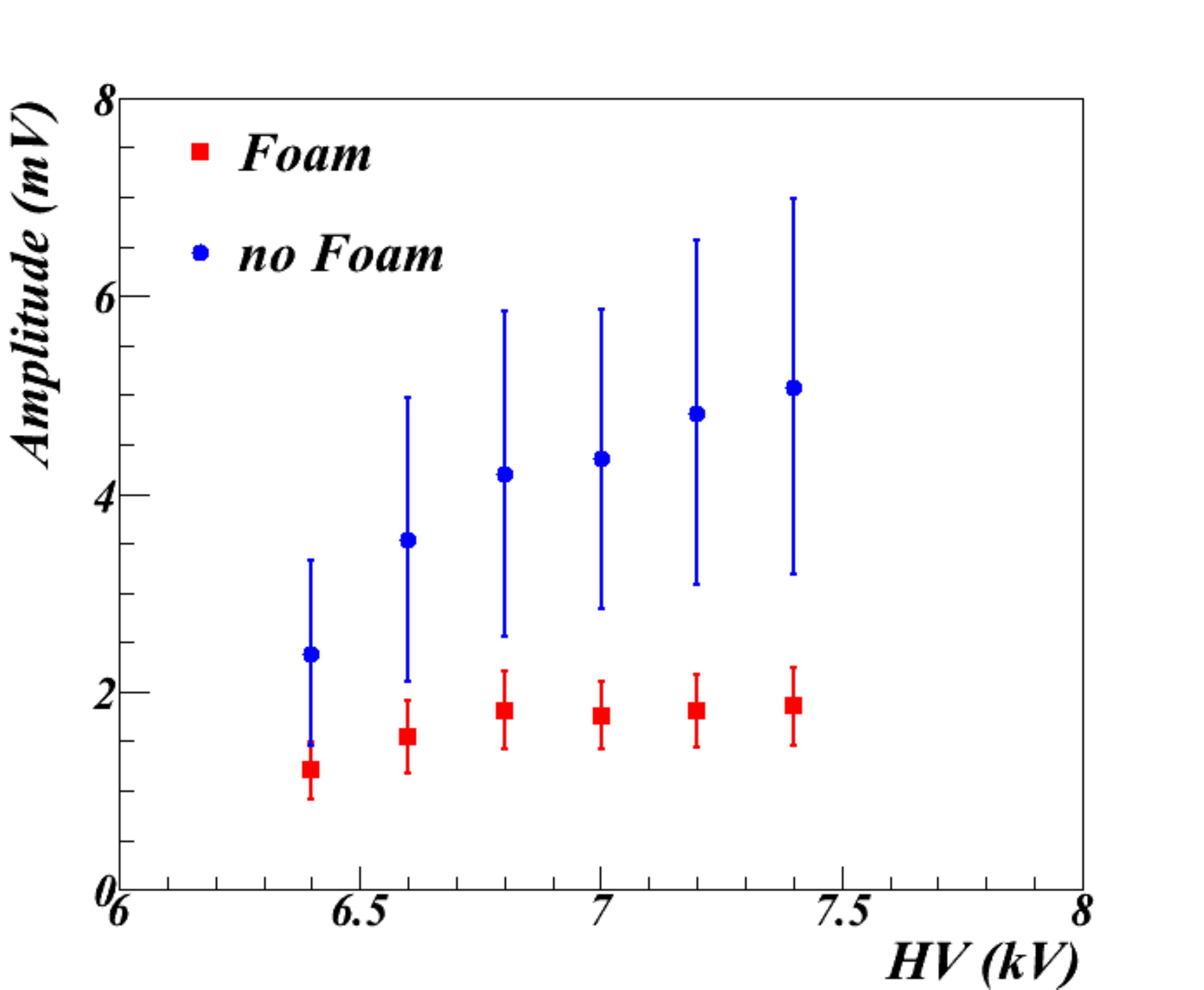}
  \end{center}
  \vspace{-0.3pc}
   \caption{Mean Amplitude versus applied HV, with/without foam, as measured at the experimental site of YangBaJing.}
   \label{MeanPeak}
\end{figure}
As reported in \cite{aielli06}, a foam layer  of a few millimiters has been added between the RPC  and the BP in order to avoid the coupling with the strip plane mounted on the other side of the RPC. In Fig\ref{MeanPeak} is reported the mean amplitude of the m.i.p. signal versus the applied HV, with and without the foam layer. It can be seen that in case of no foam there is a much faster increase of the measured amplitude, which is higher by a factor 2$\div$3 in the efficiency plateau; the foam instead stabilizes the signal and makes the plateau extension more evident.  Similar behavior has been observed for some parameters like strip multiplicity and pulse height of the strips with respect to the thickness of the foam layer \cite{aielli06}.

\section{Intrinsic linearity of the RPC}
In order to check the intrinsic linearity of the RPC response up to particle densities of a few 10$^4$/m$^2$, which is the density at the core of showers induced by primary cosmic rays of  PeV energies, a test-beam on small RPCs has been performed in October 2009 at the Beam Test Facility (BTF)\cite{BTF03}, part of the DAFNE-factory complex (INFN National Laboratory of Frascati, Italy). The facility includes a high current electron and positron LINAC, a 510 MeV e$^-$ and e$^+$ accumulator and two 510 MeV storage rings. The BTF is a beam transfer line optimized for the production of electron or positron bunches, in a wide range of multiplicities down to single-electron mode, in the energy range between 50 and 800 MeV. The typical pulse duration is 10 ns, very close to the time thickness of the shower front near the core,  and the maximum repetition rate is 50 Hz.

\begin{figure}[t]
  \begin{center}
    \includegraphics[width=4.0in]{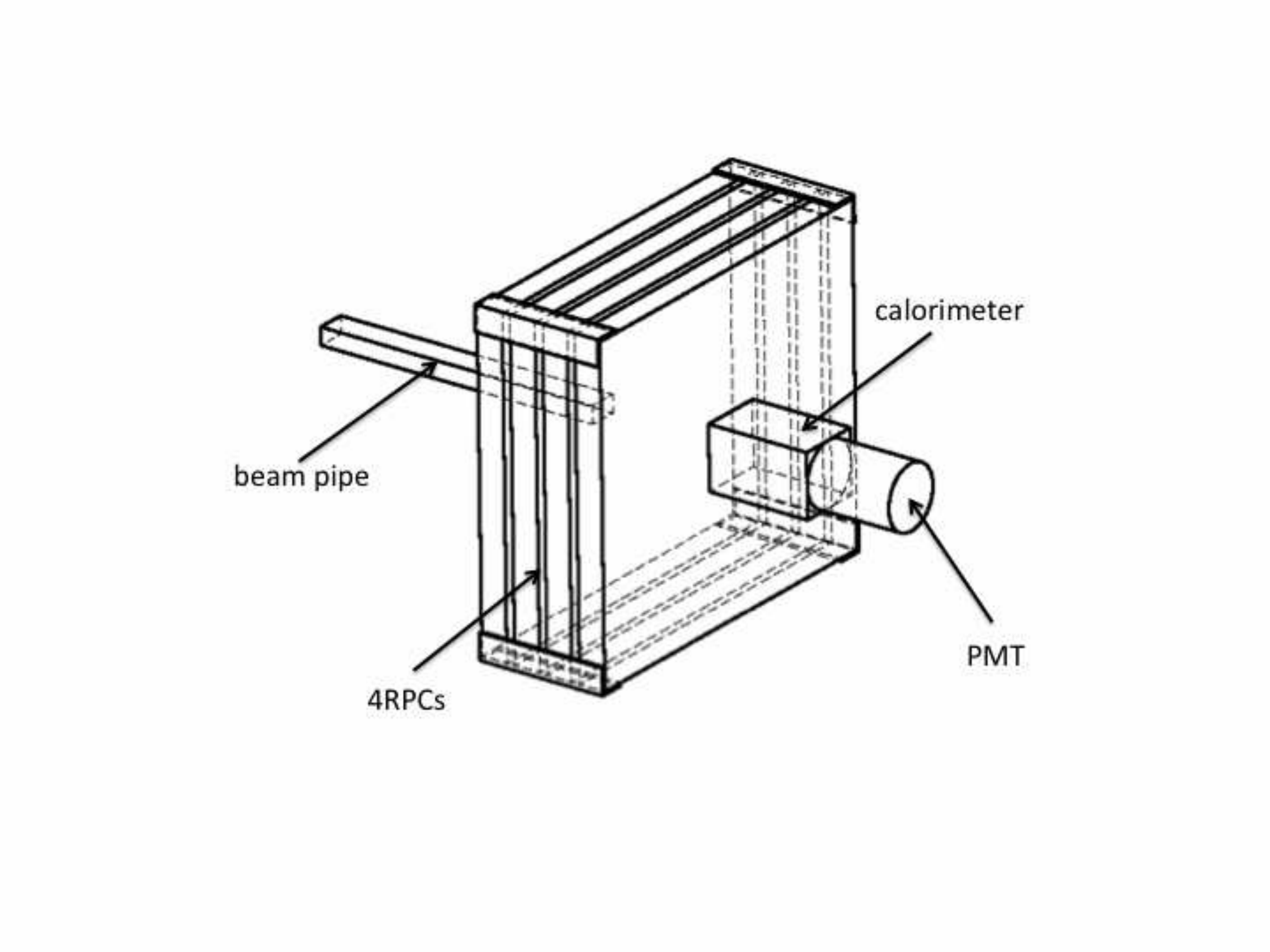}
  \end{center}
  \vspace{-0.3pc}
   \caption{Experimental setup of the test beam at the BTF.}
   \label{BTF_Assonometria}
\end{figure}

\begin{figure}[t]
  \begin{center}
    \includegraphics[width=4.0in]{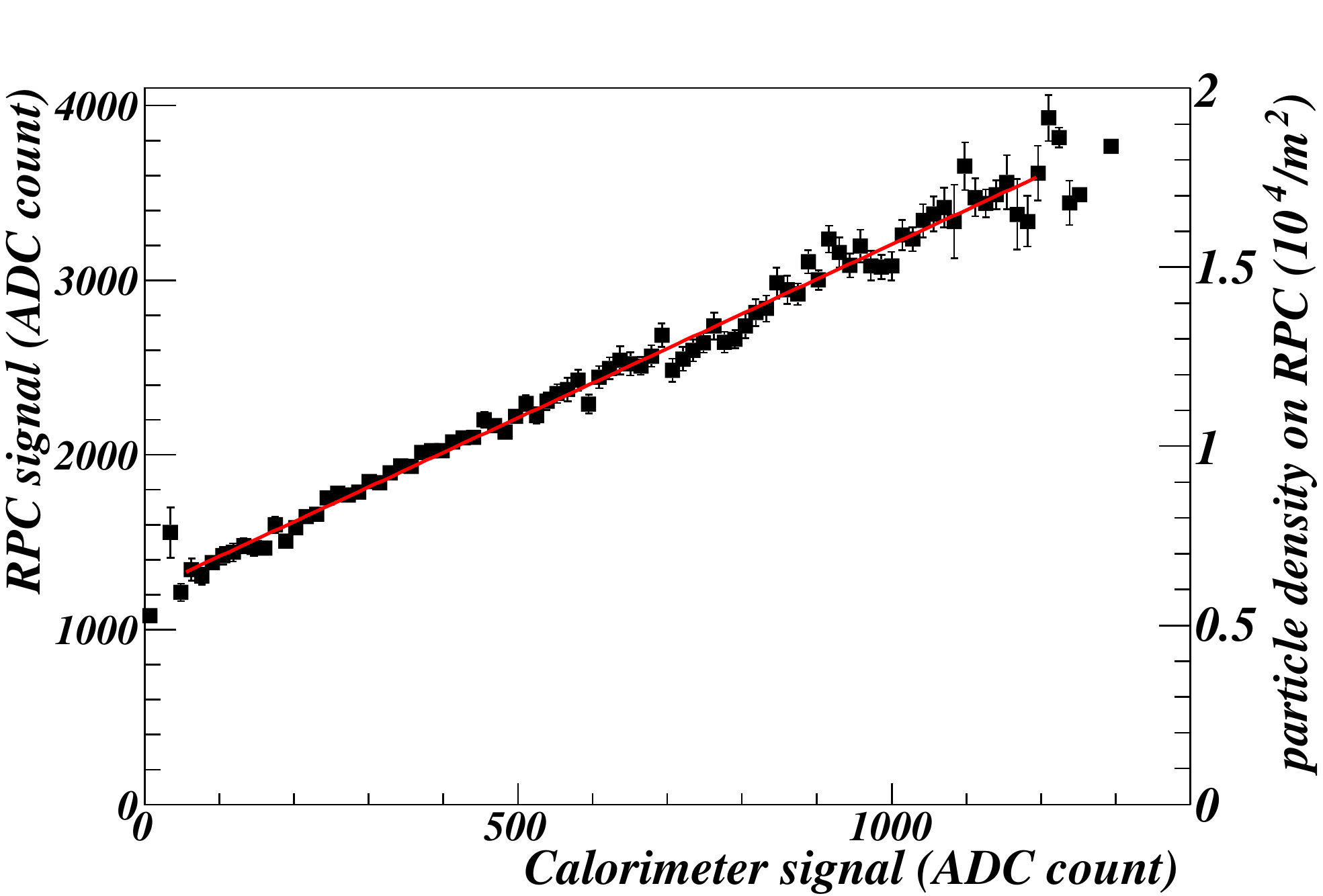}
  \end{center}
  \vspace{-0.3pc}
   \caption{Result of the RPC linearity test performed at the BTF (see text for details).  The fit with a straight line, in red, has been performed.}
   \label{BTF_Result}
\end{figure}

\begin{figure}[t]
 \begin{center}
    \includegraphics[width=4.0in]{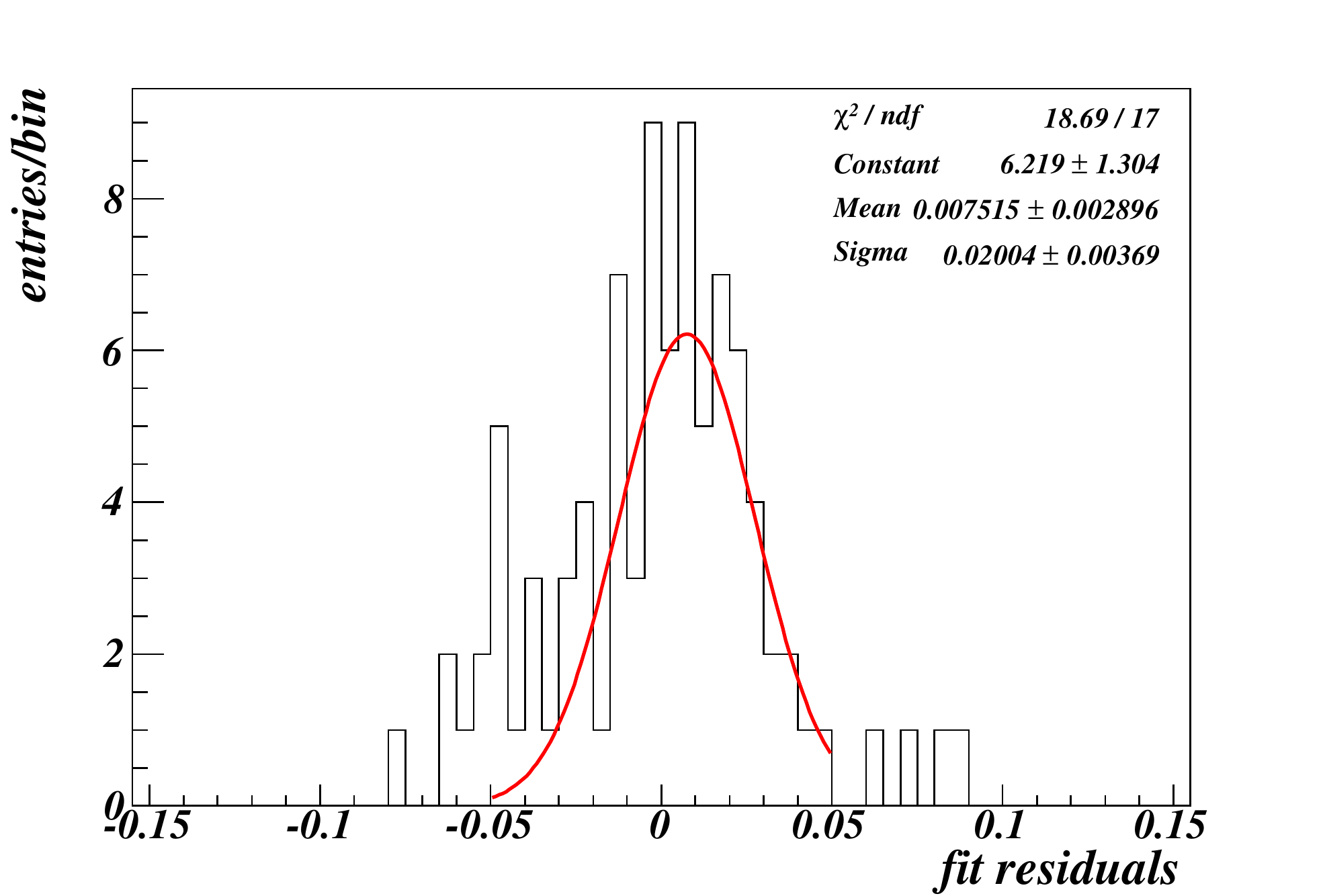}
   \end{center}
   \vspace{-0.3pc}
   \caption{Residuals normalized to the fit values (see Fig.\ref{BTF_Result}). The gaussian fit to the distribution shows a good agreement (see text for details).}
   \label{Fit Residuals}
\end{figure}

The experimental setup of the test at BTF is shown in Fig.\ref{BTF_Assonometria}.
Four RPCs of dimensions $\rm{63\times57}$ cm$^2$, having the same layout as the standard ones (see \cite{aielli06} and Fig.\ref{RPCLayout}), in particular provided with 8 strips for digital readout on one side of the gas volume and one pad for analog readout on the other side, were put just at the exit of the beam line. With all transfer line quadrupoles off during the test, the beam spot was completely de-focused and limited, at the exit,  by the vacuum pipe section which has dimensions $\rm{5\times3}$ cm$^2$. The bunch rate was  1 Hz and the slits defining the beam characteristics were operated so to have a number of particles ranging from a few up to many tens per bunch. The energy was in the range 450-500 MeV. The RPCs were operated with the same mixture used in the ARGO-YBJ experiment and a check of their performance was done before the test beam by stacking the 4 RPCs in a vertical telescope, then measuring  the efficiency and the operation voltage with cosmic muons crossing the telescope. Above the knee voltage, $\sim$ 9.2 kV,  in the plateau region, the efficiency was $\ge$ 95$\%$. Both values were in agreement with the expectations.
At  9.2 kV, the voltage of operation during the test, the measured amplitude of the pad signal for a single particle crossing the detector was 3.7 $\pm$ 0.1 mV. A small lead glass calorimeter, located just beyond the RPCs (see Fig.\ref{BTF_Assonometria}) and aligned with its major axis along the beam direction, was used to measure the particles exiting the RPCs. The lead glass block is from the former OPAL experiment \cite{OPALexp} and has the shape of a truncated prism of Schott SF57 lead glass; the block length is 37 cm, corresponding to 24 radiation lengths, with a base of  $\rm{11\times11}$ cm$^2$. The light signal is read by  a Hamamatsu R2238 photomultiplier.
The data acquisition was performed by a custom system managed by a Motorola 6100 CPU and  housed in a VME crate.
The calorimeter signal was used for cross comparison with the RPC signals, particularly with the signal of the first RPC along the beam line. It was acquired, after suitable adaptation, by means of the same electronics used for the RPC analog signals. The trigger  to the DAQ was provided  by the signal of bunch-crossing, available through the BTF system. The beam geometry (refer to \cite{BTF03} for details), even with the beam fully de-focused,  guaranteed an  almost complete containment of the beam by the calorimeter, or very marginal loss of information (the Moliere radius of the Schott SF57 lead glass is 2.61 cm).
The scatter plot of  Fig.\ref{BTF_Result} shows the signal of the first RPC along the beam line (mean value $\pm$ r.m.s.) in bins of ADC counts in the calorimeter. Assuming the linearity of the RPC behavior and the amplitude of the single particle mentioned before, we reconstructed between 7 to 30 particles impinging on the RPC surface. This number is fully consistent with the estimate provided by the beam monitoring system of the BTF. On the right scale of Fig.\ref{BTF_Result} the corresponding particle density of the beam is reported.
To check the consistency with linear response, the experimental data have been fitted with the red straight line shown in Fig.\ref{BTF_Result} and the residual values, normalized to the fit values, are reported in the histogram of Fig.\ref{Fit Residuals}. The gaussian fit to the residual distribution (Fig.\ref{Fit Residuals}) shows a good agreement, as confirmed by the value of the $\chi^2/d.o.f.$. From the fitted values of the gaussian parameters one can say that local deviations are contained within a few per cent (r.m.s) , while the integral deviation (mean) is below 1$\%$.
The offset of the RPC response in Fig.\ref{BTF_Result} is due to the strong attenuation of the calorimeter signal and to its adaptation to match the specifications of the readout electronics. In conclusion,  up to 30 particles on 15 cm$^2$ there is no evidence of deviation from linearity behavior of the RPC, which means linearity response up to density of about $\rm{2 \times 10^4/m^2}$.
Of course this value is conservative because the particle density of the beam spot is not properly uniform.

\section{Local Station and Trigger System}
\begin{figure*}[th!]
\centering
\includegraphics[width=5.0in]{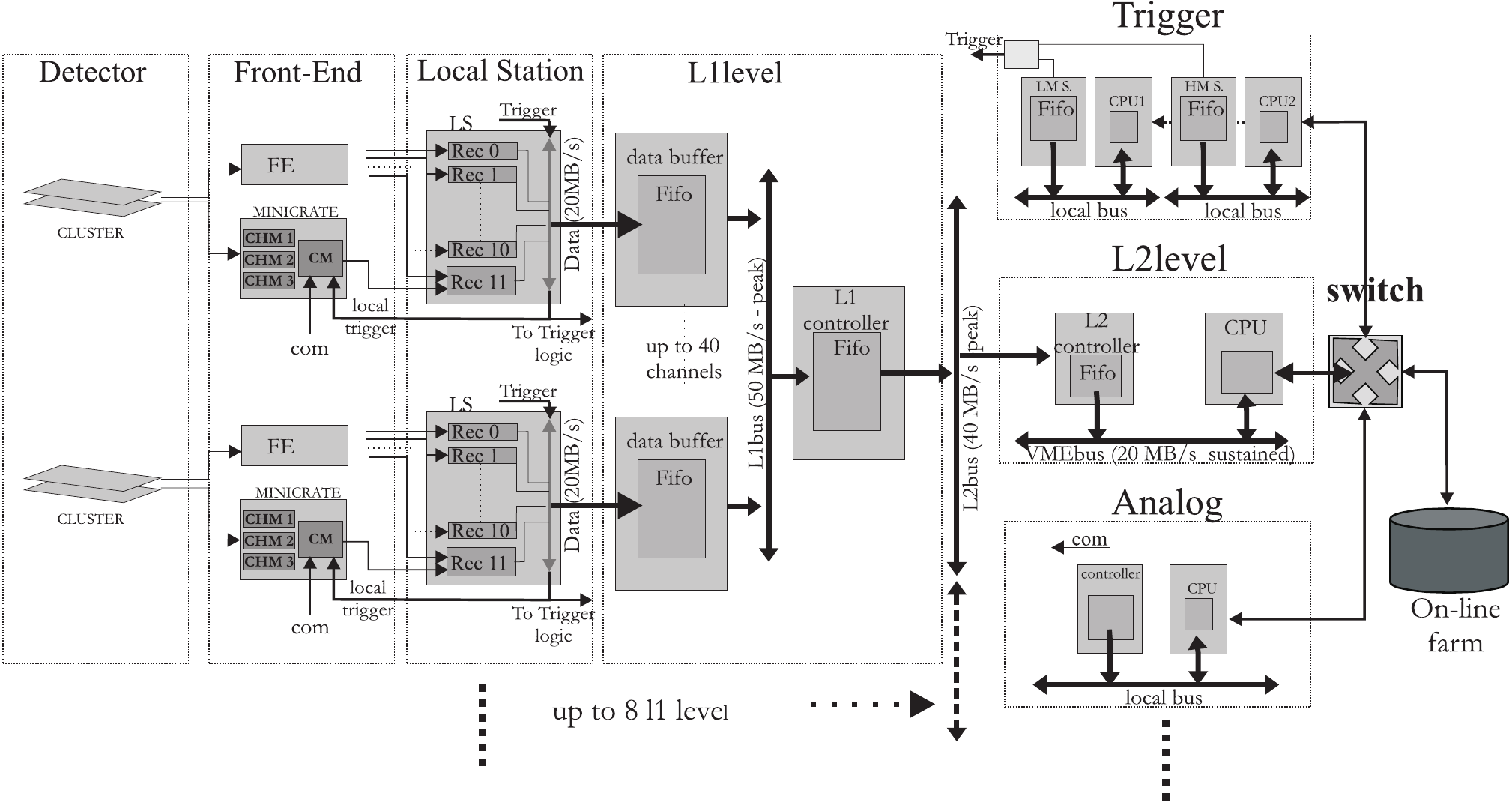}
\caption{Schematic drawing of the Front-End electronics and DAQ system. The FE block is based on a 8-channel discriminator chip for strip signal processing located inside the metallic casing of the RPC. The  MINICRATE block is in charge of the BP signal processing. The MINICRATE electronics and the LS crate are located in the middle of the Cluster unit; each LS in turn is connected to the Central Station with a star-like custom network for triggering and data transfer purposes.}
\label{daq_sch}
\end{figure*}

The trigger of the experiment is generated by the digital signals sent by the Front-End boards mounted on the RPCs. These digital signals are processed by a specific crate named Local Station (LS) \cite{assiro} - the Cluster DAQ Unit -,  as depicted in Fig. \ref{daq_sch}, that provides the pad multiplicity to the trigger system.  The LS crate contains and manages 12 receiver cards, one I/O card for the communication with the DAQ and one active backplane.
Each receiver card collects the signals coming from one RPC chamber and provides the fast-OR signals which start the TDC counting. When a trigger occurs, a common stop signal goes from the backplane to the receiver cards, which store the patterns of the active strips, and to the TDCs which perform the arrival time measurement.
Each LS outputs two busses, namely a 6-bit Low Multiplicity (LM) weighted bus (providing signal when $\ge$1, $\ge$2, $\ge$3, $\ge$4, $\ge$5, $\ge$6 pads are fired within 150 ns) and a 4-bit High Multiplicity (HM) weighted bus (providing signal when  $\ge$7, $\ge$16, $\ge$32, $\ge$64 pads are fired within 60 ns).\\
The trigger system \cite{trigger}, which has the LM and HM busses in input, implements two different selection algorithms based on a simple, yet robust, majority logic which takes into account the topology and the time distribution of the fired pads. The LM trigger implements a selection of small-size showers by requiring at least $N_{trig}$ fired pads in the central carpet. The HM trigger has been designed to select showers with a much higher particle density. The whole trigger electronics is hosted in 2 VME crates as shown in Fig. \ref{daq_sch}.\\
The LM  trigger is implemented with a four-level hierarchical architecture, where each level correlates only pads belonging to adjacent areas. According to simulations and to the measured pad rate ($\sim$400 Hz/pad), the number of spurious hits in the 420 ns trigger window has been estimated to be less than 3.\\
The data collected in each LS, that is the pattern of the fired strips and the arrival time of the particles, are packed and transmitted at each trigger occurrence to the central DAQ at a rate of 160 Mbit/s (16-bit word in 100 ns) by means of the I/O card.\\
The present trigger set-up enables just the LM selection with a threshold of 20 pads. Since the amount of data for each event strongly depends on the shower size and the cosmic ray spectrum follows a power law, the data frame of the event ranges from about hundred bytes to Mbytes, with an average event size of about 2 kbytes.

\section{The Charge Readout System}
\begin{figure*}[th!]
\centering
\includegraphics[width=5.0in]{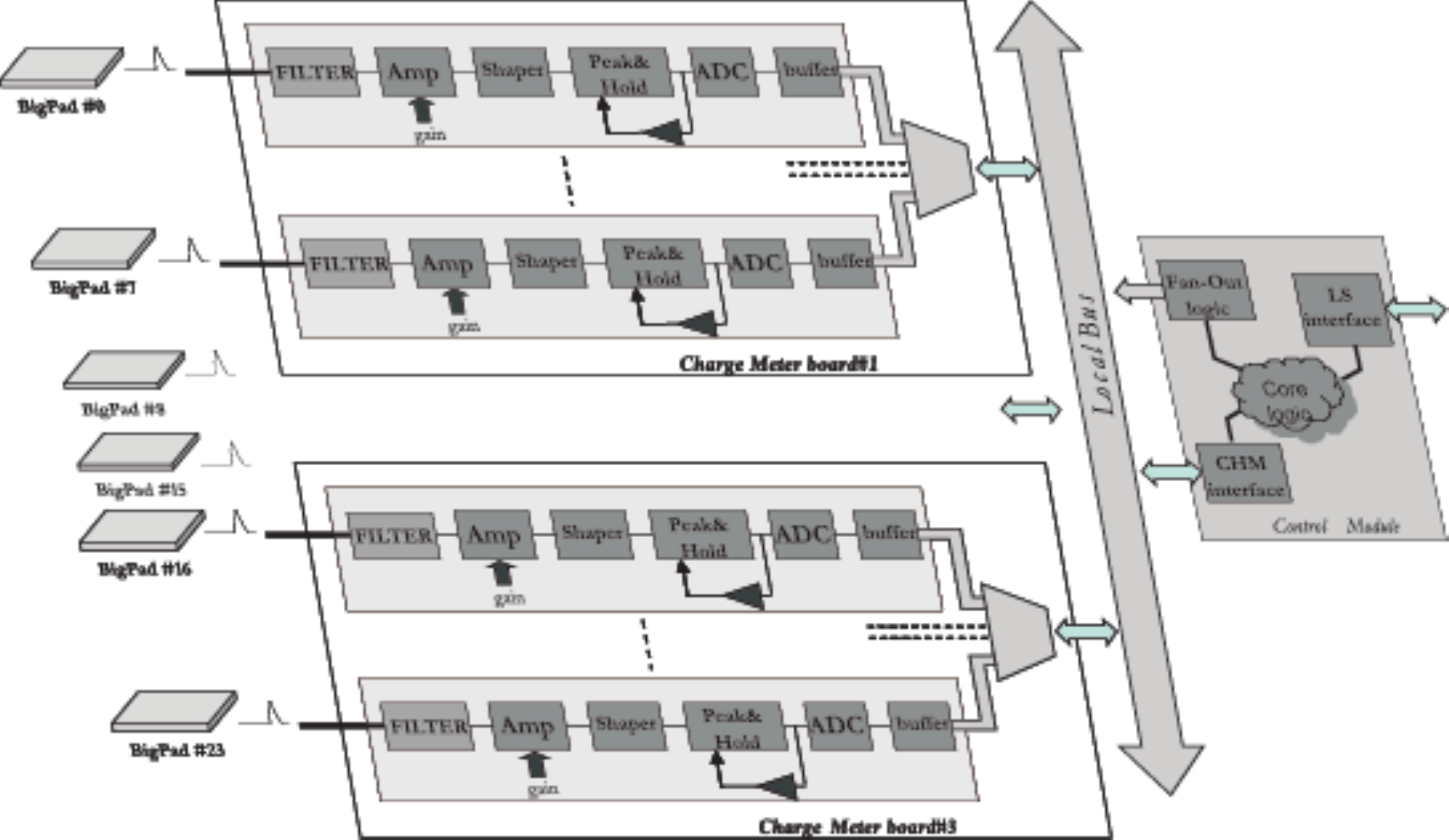}
\caption{The logic block diagram of a MINICRATE section.}
\label{minicrate_sch}
\end{figure*}

The BP signals of two adjacent Clusters are processed by electronic modules hosted in a custom crate, called MINICRATE, that has two independent sections, each one containing 3 readout cards (CHargeMeter cards) and a Control Module (see Fig.\ref{minicrate_sch}). The CHargeMeter (CHM) processes 8 analog signals and digitizes them, while the Control Module builds the data frame of 3 CHM boards and transfers it to the LS, which finally provides the data to the central DAQ. A 12.5 m coaxial cable is used to feed the signal to the CHM input adapted to 50 $\Omega$. The CHM board is made of 8 identical sections. The first block of each section consists of a voltage limiter able to cut out spikes greater than 40 V to protect the next electronic stages.
It is followed by a linear amplifier with a dynamic gain set by a 3-bit programmable register, which provides 8 input values for full scale (f.s.) setting, as show in Tab.\ref{fsValues}.
\begin{table}
\begin{center}
\begin{tabular}{|c||c|c|c|c|c|c|c|c||}
\hline \hline
\small{ reg. value ($\rm{R_{fs}}$)}  & \small{7}& \small{6}& \small{5}& \small{4}& \small{3}& \small{2}& \small{1} & \small{0}\\
\hline
\small{ f.s. voltage (V)} & \small{0.29} & \small{0.58}& \small{1.16}& \small{2.13}& \small{4.26}& \small{8.52}& \small{16.2}& \small{32.4}\\
\hline \hline
\end{tabular}
\end{center}
\caption {\small{3-bit register values and corresponding effective full scale (f.s.) voltages; each f.s. voltage has a spread of 2$\div$2.5$\%$} due to the different board productions. The scale of operation is indicated by G$n$ with $n$ equal to the value of $\rm{R_{fs}}$.}
\label{fsValues}
\end{table}
Operating the system at different scales allows an overlap between digital and analog readout which is an efficient way to calibrate the analog system. The scale of operation is indicated by G$n$ with $n$ equal to the value of $\rm{R_{fs}}$  as shown in Tab.\ref{fsValues}.\\
The sensitivity of each scale is about 0.2$\%$ of the full scale value, below this value the electronics sensitivity is reduced.\\
The signal at the output of the amplifier ranges from  0 to 2.5 V which is the reference input of the Analog to Digital Converter (ADC).\\
In the shaper module the long tail of the BP signal is cut out by doubling and inverting the signal, then summing the original to the inverted one with a $\Delta t$ delay. This time delay has been set to about 500 ns mainly to guarantee that the analog signal reaches its maximum and, secondly, to be sensitive to delayed particles in the shower front. A Peak and Hold ($P\&H$) circuit represents the core of the next stage. It continuously  samples the output of the previous stage and keeps the highest reached value for 2 $\mu s$; the voltage drop per time unit of the $P\&H$ is $\sim$ 4 mV/$\mu s$, which does not cause any trouble with respect to the trigger jitter time as already shown in \cite{iacovacci03}.
If a conversion signal  (named local trigger) arrives within 2 $\mu s$, the ADC starts to digitize the $P\&H$ amplitude.
The conversion signal is generated by the HM bus of each LS and can be selected with different thresholds (namely $\ge 16, ~\ge 32~, ~\ge 64~$  and $\ge 73$ fired pads).\\
If no local trigger arrives within 2 $\mu s$, the $P\&H$ is reset and then it starts again to sample the signal from the previous stage.\\
The ADC data collection is managed in each section by the Control Module via a custom bus protocol operated on the backplane lines. The Control Module receives the local trigger from the LS and distributes it to the 3 CHM units starting the ADC digitization. Then, each CHM board replies by asserting a local Busy signal in order to prevent the generation of further local triggers  while the ADC is converting. After 14 clock cycles at 10 MHz the data are converted, latched and ready to be collected by the Control Module.\\
In this project the AD7472 component \cite{ADC} by Analog Devices has been used. It is a 12 bit high-speed ADC offering an high throughput with a low power consumption (1.5 MSPS with 4 mW power consumption). The differential nonlinearity (DNL) measured by the producer is 1.9 LSB.
Some tests about performance, homogeneity of the board production, homogeneity of the electronic channels, linearity and calibration features have already been done and results have been reported in \cite{iacovacci03}, \cite{iacovacci05}, \cite{iacovacci09}, \cite{iacovacci2010} and \cite{sheng09}.\\

\section{The Data Flow}
In this section we describe the analog data flow, from the FE electronics up to the event builder, paying attention to the synchronization with the digital frame and the overall system.\\
After the digitization of the BP signals, the data are sent to the FIFO of a special receiver board (Rec11 in Fig. \ref{daq_sch}) hosted in the LS. The transfer to the central DAQ starts when the experiment trigger confirms the local trigger.
A multiplexer logic provides a simple and flexible data readout. Each CHM board readout is performed in 4 clock cycles (400 ns, thus the whole information is transferred in 1.2 $\mu s$ to the FIFO of Rec11 in the LS.
If the local trigger is not validated by the experiment trigger, the data stored in Rec11 are discarded and the ADC system will be ready to process a new local trigger. In both the cases the Busy signals generated by CHM boards are reset.

When the experiment trigger arrives at the LS, the event building process starts and the event number, the addresses of the fired strips, the pad time information and the BP charge information are packed in a data frame which is transmitted  to the DAQ system, then stored into the FIFO memory of the Data Buffer board \cite{Amb} of the Level-1 (L1) readout system (\cite{daq}, \cite{icrc01}). A simplified block diagram of the data flow is shown in Fig. \ref{daq_sch}. \\

The ARGO-YBJ DAQ modular structure allows a high speed and efficient data collection. It is built on a two-layer read-out architecture implementing an event-driven data collection by using two custom bus protocols, based on VME-bus \cite{VMEbus}. A Motorola VME Processor MVME6100, is responsible for the complete read-out of a Level 2 chain.
Three MVME6100 CPUs are used to manage and control the trigger and the charge readout system.\\
In the last part of this section we show the result of a test on the data collection performance with different analog configurations.
\begin{figure}[!t]
\centering
\includegraphics[width=4.0in]{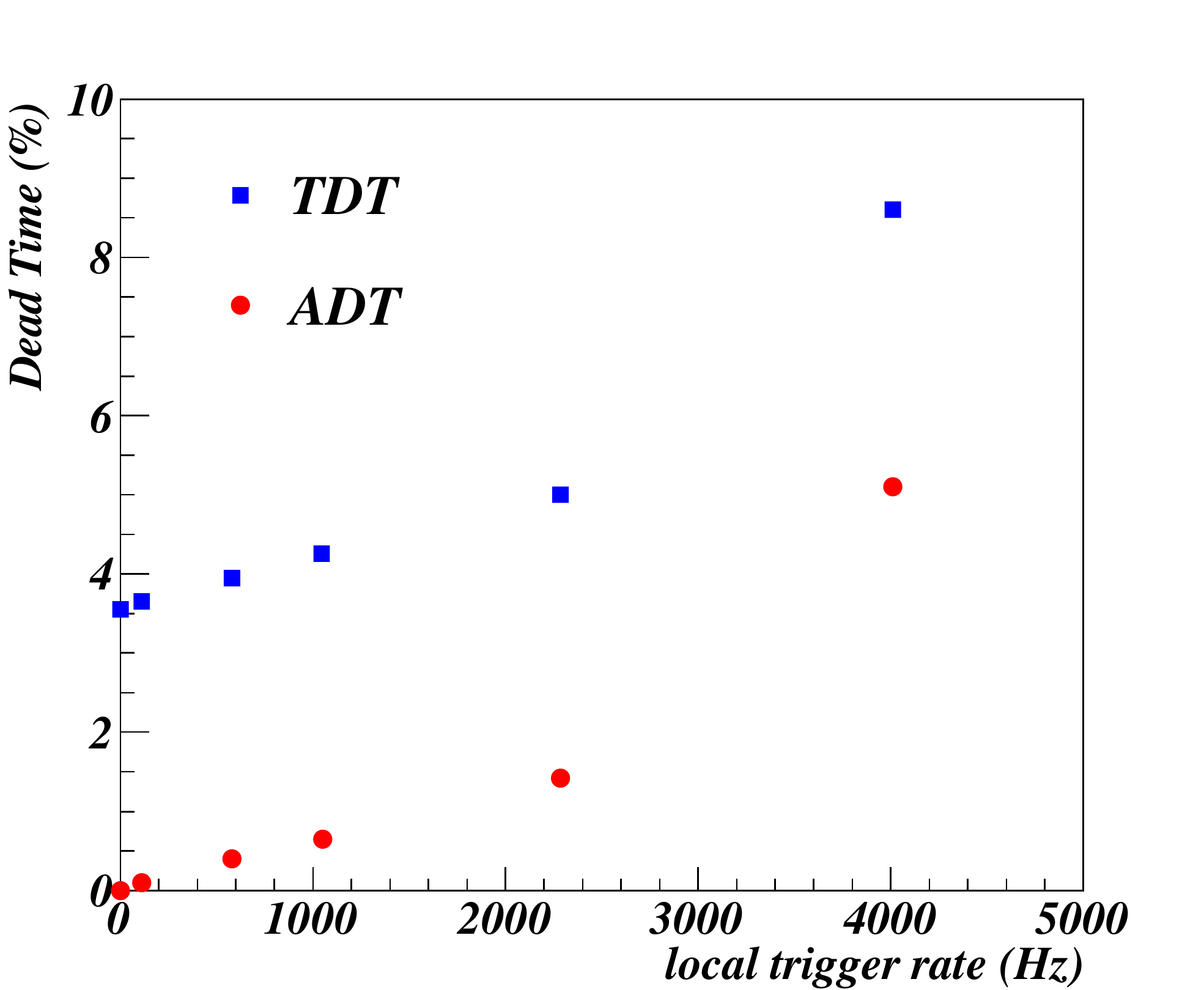}
\caption{Average values of Total Dead Time (TDT) and Analog Dead Time (ADT) versus the local trigger rate.}
\label{dt_vs_loct}
\end{figure}
Fig. \ref{dt_vs_loct} shows the behaviours of the Analog Dead Time (ADT) and the Total Dead Time (TDT)  as a function of the local trigger rate.  The fraction of the TDT due to the ADT is less than 10$\%$ for a local trigger rate of 1kHz, then it increases up to 60$\%$ at 4 kHz. As the local trigger rate for the ARGO-YBJ detector does not exceed 100 Hz, it is clear that the analog readout does not introduce any significant dead time; we estimate the ADT to be less than 2$\%$ of the TDT at the experimental conditions.

\section{The Run Control}

The Run Control has been developed in order to provide the interface between the shift operator and the DAQ components for managing and monitoring the run. The Run Control is based on the following modules: the run manager which allows the operator to startup a new run; the log of the inter-processes communicator that provides messages coming from the DAQ nodes; the run monitoring that shows some run information like run number, trigger issued, time left, dead time etc. The Run Control system has been successfully used to control the DAQ system.\\
The commands pertaining to the charge readout system, issued by the Run Control both for hardware setup and monitoring/debugging tasks, are distributed to the Analog VME CPU as shown in Fig. \ref{daq_sch}. All the MINICRATEs are connected in a daisy-chain network to the controller boards hosted in the VME crate allowing a two-way serial communication line at the speed of 1 Mbit/s. The processes pertaining to the CHM boards are managed by the Control Module which sets the ADC full scale of operation and the local multiplicity threshold for triggering.  \\
Since the CHM boards, located near each Cluster, are distributed over a very wide area they could suffer from variations of the environmental parameters, therefore their calibration and continuous control are fundamental. The calibration runs are started by the Run Control: a full scale is set, then signals with known amplitude are generated and sent to the ADC inputs, finally the digitized  values are collected and flagged properly. A Digital-to-Analog Converter (DAC902 \cite{DAC}) from Burr-Brown Products by Texas Instruments, hosted in the Control Module, generates pulses of variable amplitudes.  When a calibration pulse is generated, a fake HM signal is also generated in the LS at same time. This signal is processed by the trigger system which provides the Common Stop signal that starts the readout on the CHM board. The DAC902 is a high-speed DAC offering a 12-bit resolution. To improve the signal/noise ratio, the differential configuration output in conjunction with the operational amplifier has been used. The characterization of the DAC device shows a linear response in the whole range of 12 bits. Its typical differential nonlinearity (DNL) is 0.5 LSB.\\

\section{{Calibration and Stability of the Detector }}
In this section we discuss both calibration and stability of the RPC analog detector. The detector performance is shown, the steps of the calibration procedure are described and the related results discussed.

\subsection{The analog data taking}
The analog system has been completely assembled at YangBaJing in its final layout on the 130 central Clusters and then put in data taking since December 2009.

\begin{figure}[h]
\begin{center}
\includegraphics[width=4.0in]{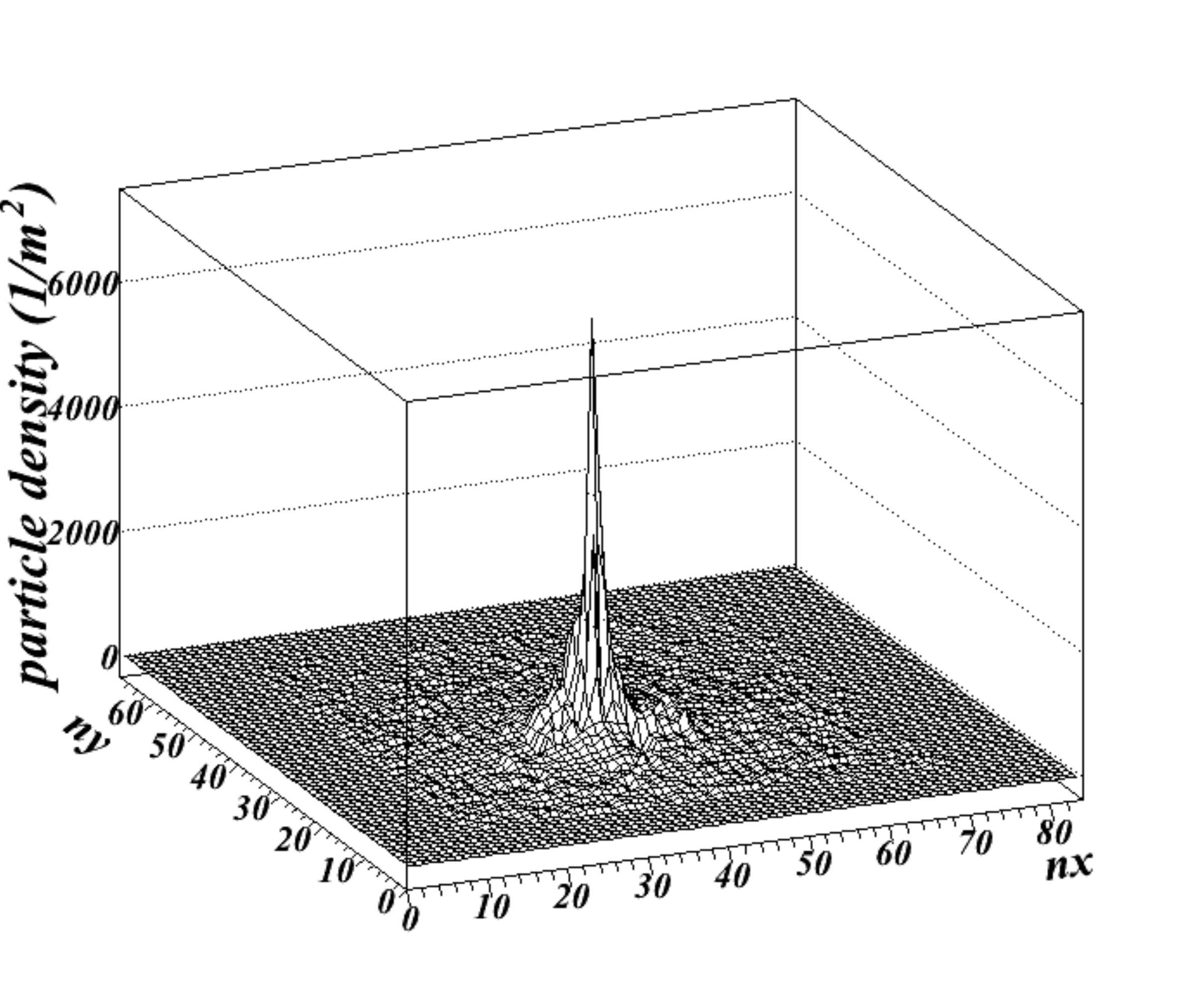}
\end{center}
\caption {Shower as recorded by ARGO-YBJ through the analog readout system. According to Montecarlo simulations, the shower would correspond to a primary proton of about 2 PeV. In the vertical scale is reported the particle density measured in the BigPad, whose area is about 1.7 m$^2$;
nx and ny are the coordinate of the BigPad. The area with low particle counts corresponds to the central carpet($\sim $78$\times $74 m$^{2}$.)}
\label{fig Analog_Showers}
\end{figure}
The system has been operated at f.s. 0.29 V ($\rm{R_{fs}}$=7, maximum gain) till June 2010. At this scale the sampled particle density overlaps with the particle density measured by the digital readout system. These data  have been used both to study the detector behavior and for calibration purposes. From July to middle August 2010, the system was operated at the intermediate scale corresponding to f.s. 2.13 V ($\rm{R_{fs}}$=4); since middle August 2010 the f.s. 16.2 V ($\rm{R_{fs}}$=1) was set. The number of BPs in the central carpet is 3120, being 3$\%$ the number of dead or badly working channels. A typical air shower event with core hitting the detector is shown in Fig.\ref{fig Analog_Showers}.
In the vertical scale is reported the number of particles measured in each BigPad, whose area is about 1.7 m$^2$. According to Montecarlo simulations for proton primary, the shower has an energy of $\sim$ 2 PeV.

\begin{figure}[h]
\begin{center}
\includegraphics[width=4.0in]{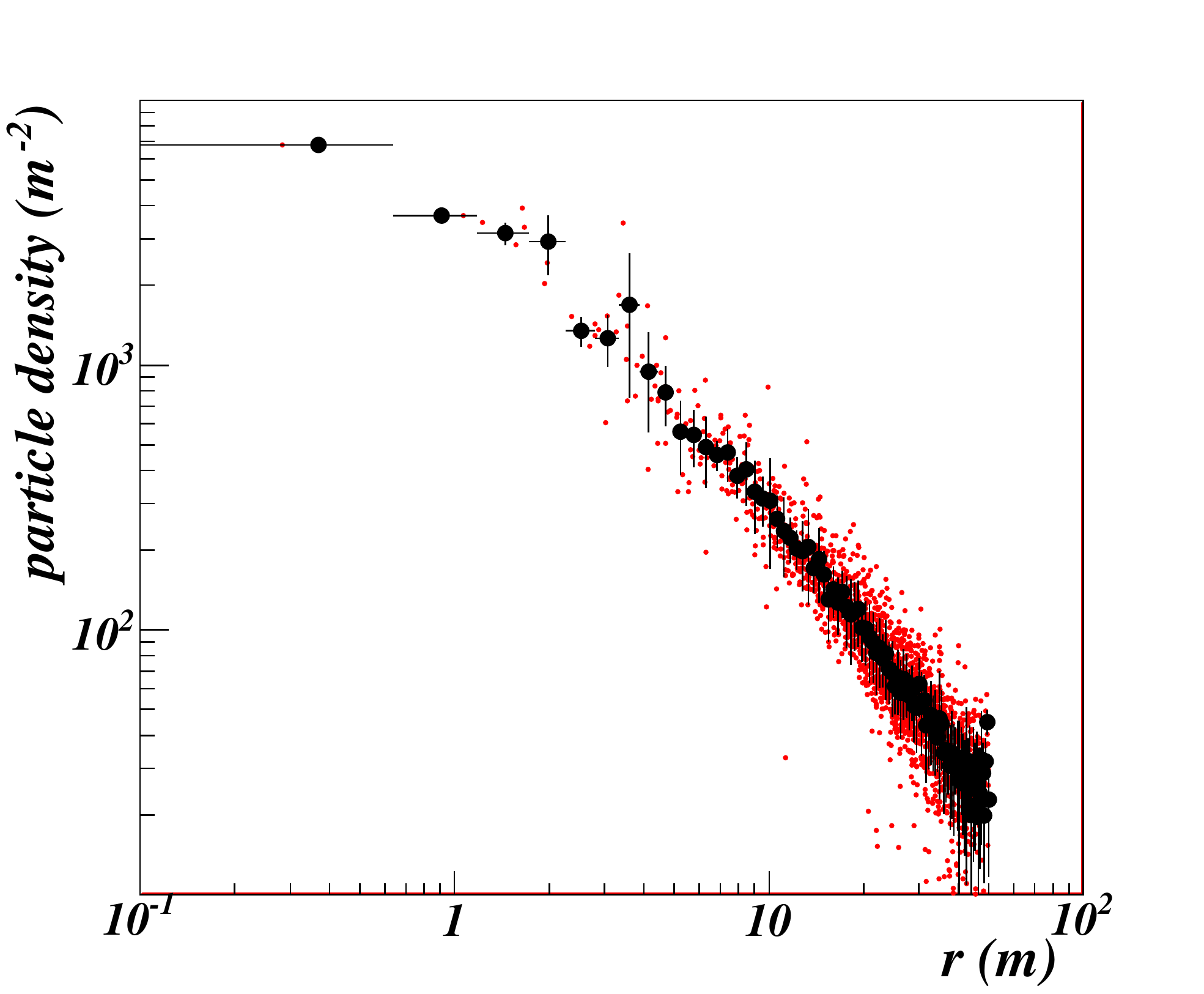}
\end{center}
\caption {Lateral distribution of the particle density for the event shown in Fig.\ref{fig Analog_Showers}. The red point are the measured density of particles on the BPs, while the black points are the mean values ($\pm$ r.m.s.) of the particle density measured in bins of distance, 0.5 m wide,  from the shower core. }
\label{LDF_shower}
\end{figure}
The lateral distribution function of this shower is reported in Fig.\ref{LDF_shower}.
Each blue point is the measured density of particles on the specific BP located at distance r from the core, which has been calculated as the center of gravity of the BP signal distribution. The black points and the their error bars correspond to mean value $\pm$ r.m.s. of the particle density measured in bins of distance from the shower core.
The figure shows the performance of the analog RPC readout: the study of the core region of high energy showers can reach an unprecedented richness of details.
Owing to the analog readout, the measurement of the proton-air cross section \cite{(pp-xsec 2009)} and the composition studies \cite{PRD_light} will be extended to PeV energies.

\subsection{\textit{The gas distribution and the BP signals}}
\begin{figure}[t]
\centering
\includegraphics[width=4.0in]{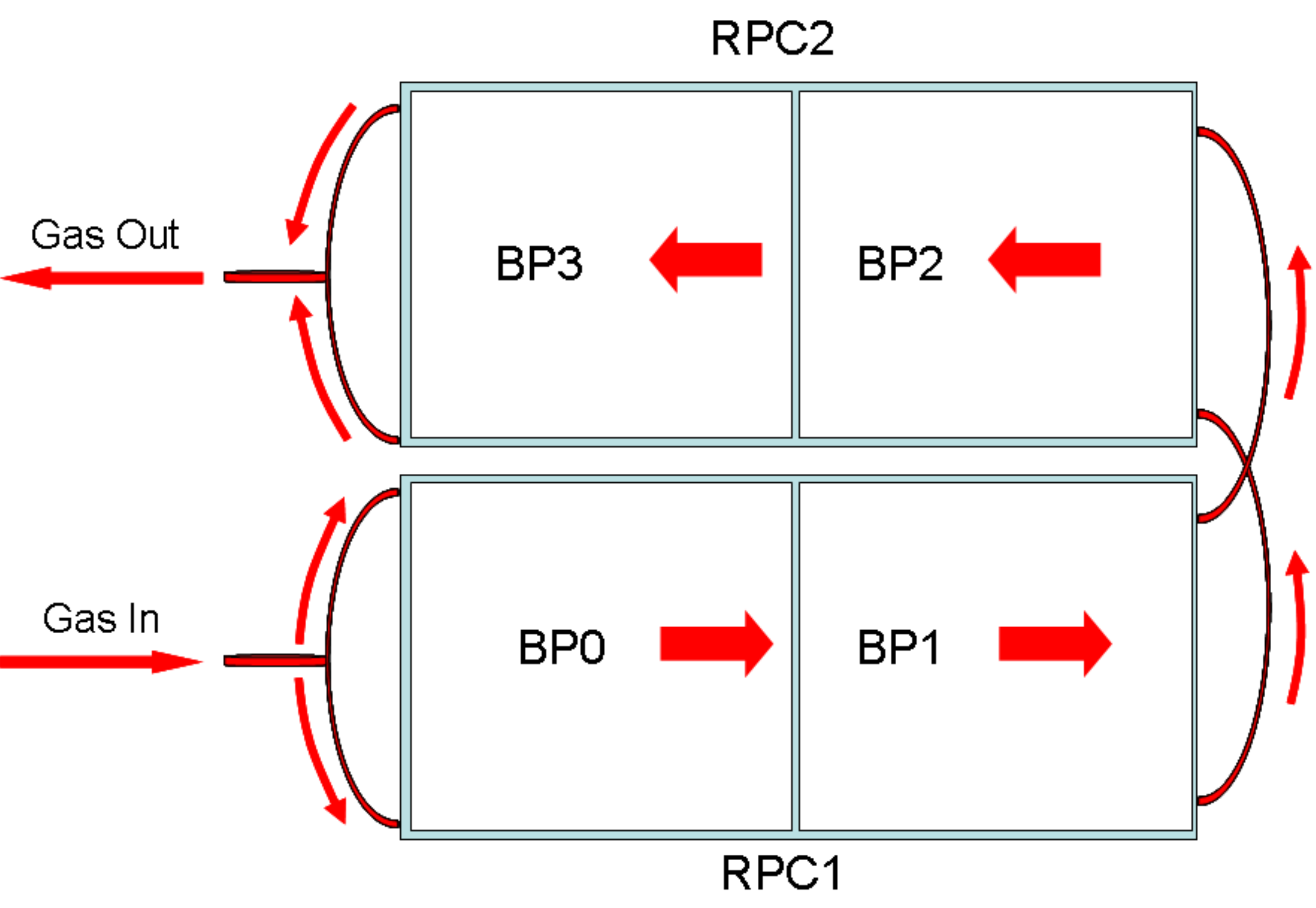}
\caption{Gas distribution: the elemental "gas channel" is shown. It consists of two RPCs serially fluxed. Inside the RPCs the gas flow is represented by the thick red arrow; the gas inlet is split in two pipes which feed the first chamber (lower left), then two pipes connect the exits of the first to the inputs of the second chamber, and finally the two output pipes are joint to form the gas outlet. Along the gas line the 4 BPs are labeled as BP0, BP1, BP2 and BP3, BP0 being the first pad and BP3 the last one receiving the gas.}
\label{gas_line}
\end{figure}

\begin{figure}[t]
\centering
\includegraphics[width=4.0in]{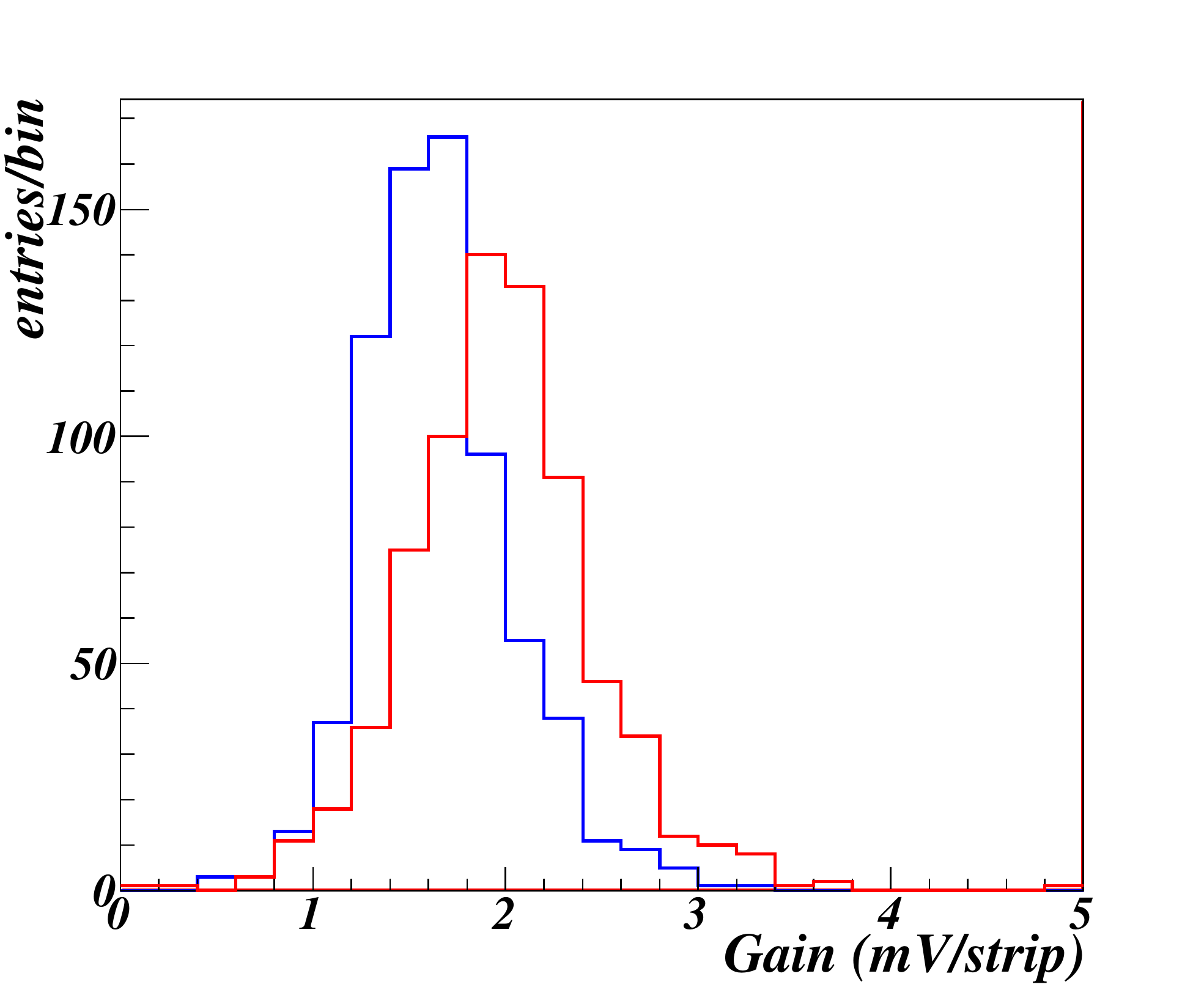}
\caption{Gain (mV/strip), or amplitude of the BP signal corresponding to one fired strip.  The gain distribution for the 1.st (BP0, red) and the 4.th (Bp3, blue) BP along the gas line is reported.}
\label{bichambergas}
\end{figure}

Up to the installation of the analog system in the ARGO-YBJ experiment,
RPCs have been mainly used for triggering and tracking purposes. On the other hand, gas detectors operated in quasi-proportional mode \cite{EAS-TOP} or in  streamer mode \cite{ALEPH} have been  successfully used in calorimeters. This is the first time RPCs are used in analog readout mode over a very large area. The gas distribution is accomplished by feeding parallel lines ('gas channels') each one made of a pair of RPCs serially connected, as shown in Fig.\ref{gas_line}. The 4  BPs along the gas line are labeled with BP0,BP1,BP2 and BP3.  A total of 780 gas channels supply the RPCs carpet while additional lines provide the gas distribution to the guard-ring RPCs.
Each gas line is equipped in input with a capillary tube (glass, 32 mm long and 0.6 mm diameter) whose impedance is so high to prevail on every other one and to define by itself the gas flow through the gas channel.
The gas volume of the 153 Clusters is 19 $\rm{m^3}$, and the gas flow is set to 3.1 volume/day; the effective gas flow, evaluated through the gas consumption, or by measuring the rate of weight decrease of the gas bottles, corresponds to 3.0 volume/day. An effect related to the gas flow has been observed, namely the amplitude decrease of the BP signal along the gas line. This effect, which is particularly strong between BP0 and BP1,  is shown in Fig.\ref{bichambergas} where the amplitude distributions are reported separately for the BP0 and BP3. The unit of measurement of the amplitude, mV/strip, assumes one fired strip to correspond to one detected particle, this point will be discusses and clarified later on (Sec.VIII.III.2). The maximum attenuation ($\sim$17$\%$ ) is observed for BP3 with respect to BP0.
Moreover, while the mean amplitudes of BP1, BP2 and BP3 are tightly correlated to the environmental parameters, specifically pressure and temperature, the amplitude of BP0 spans in a band wider than the environmental parameters allow \cite{iacovacciICRC11}. This effect was unexpected and is still matter of investigation.
\subsection{The calibration procedure}
In order to convert the ADC counts to particle number two steps are needed, namely first converting the ADC count to amplitude, then the amplitude to the number of particles. The first step is referred to as electronic calibration, the second one as gain calibration.
\subsubsection{Electronic calibration}
\begin{figure}[t]
\centering
\includegraphics[width=4.0in]{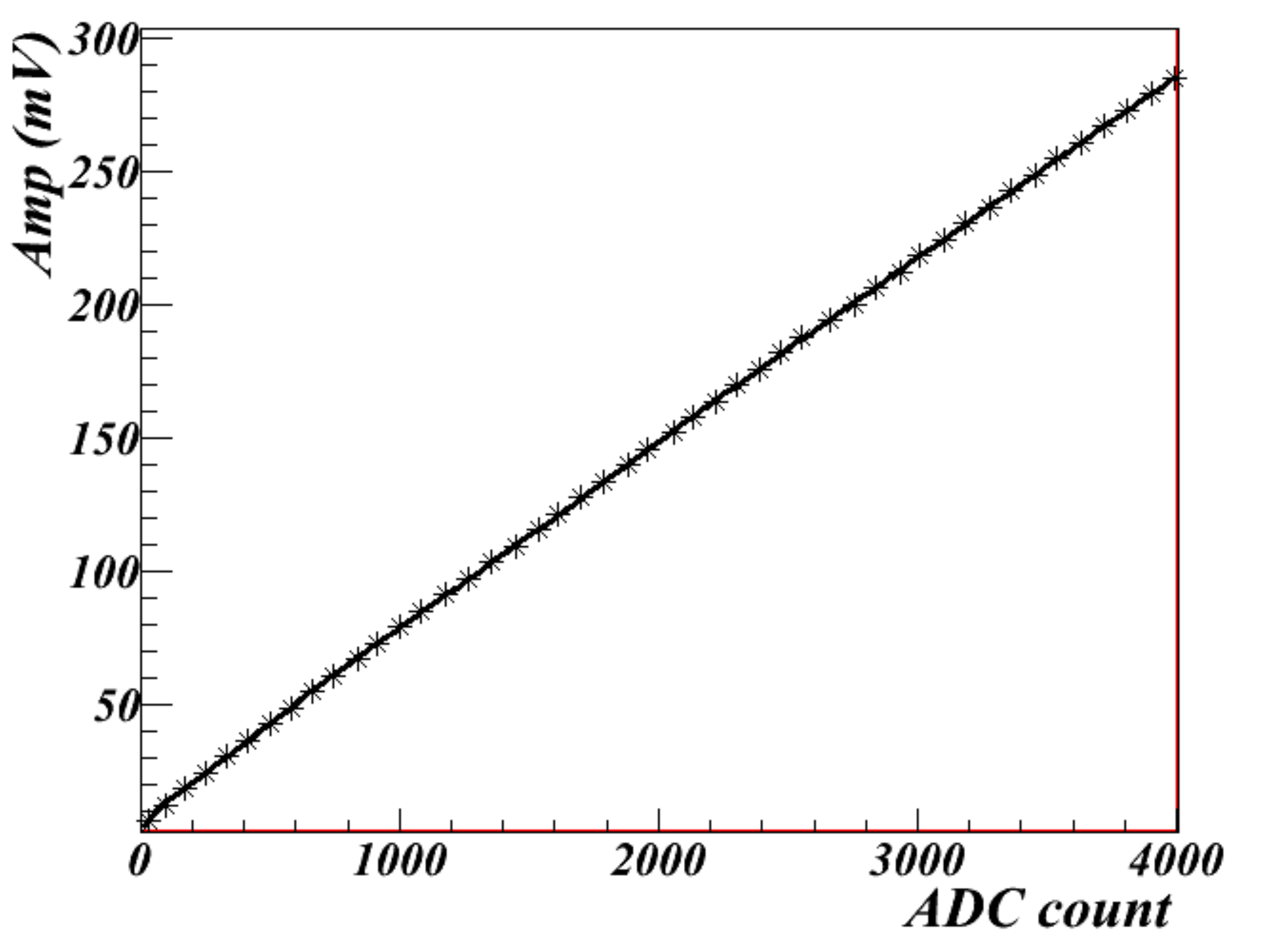}
\caption{Electronics calibration of a channel. The relation between ADC counts (output) and signal amplitude (input) of the specific channel has been obtained by fitting the experimental points (stars), provided by the calibration system (see text),  with  a suited functional form.}
\label{ElectronicCalibration}
\end{figure}

\begin{figure}[t]
\centering
\includegraphics[width=4.0in]{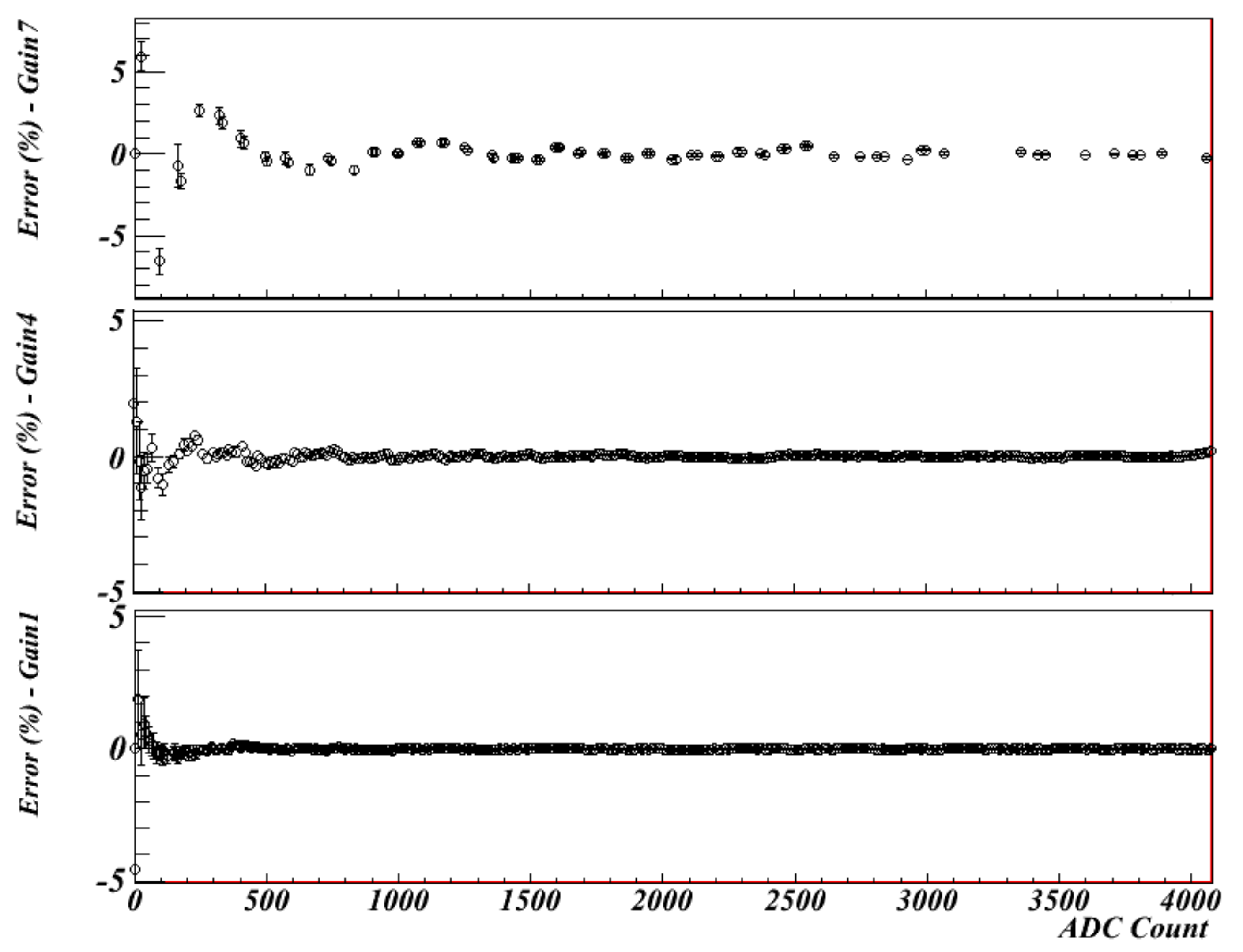}
\caption{The percent difference of the functional form with respect to the calibration points is reported versus the ADC count for 3 scales of operation.}
\label{ElectronicCalibration_error}
\end{figure}

Calibration runs \cite{iacovacci09,mastroia11} have been performed at the different f.s. the electronics has been operated;  the relation between ADC count and input amplitude, provided by the DAC, has been fitted with a suited functional form as shown in Fig.\ref{ElectronicCalibration}.  The coefficient of the linear term (P1), which is the dominant one, comes out to have a spread 1.5-2.5$\%$ depending on the board production, which in fact was achieved in three different bunches.
In Fig.\ref{ElectronicCalibration_error} the percent error of the calibration is reported versus the ADC count for the 3 scales of operation. At the f.s. 16.2 V and 2.13 V, the error is by far below 1$\%$ for ADC count greater than 100 while it is less than 2-3$\%$ below 100; at the f.s. .29 V the error is below 1$\%$ for ADC count greater than 400 and becomes 5-6$\%$ below 100, in between it is  2-3$\%$. To check the long term stability of the electronics, calibration runs have been issued several times after weeks and months;  the results showed a stability  within about 1$\%$ .\\
\subsubsection{Gain calibration}
\begin{figure}[t]
\centering
\includegraphics[width=4.0in]{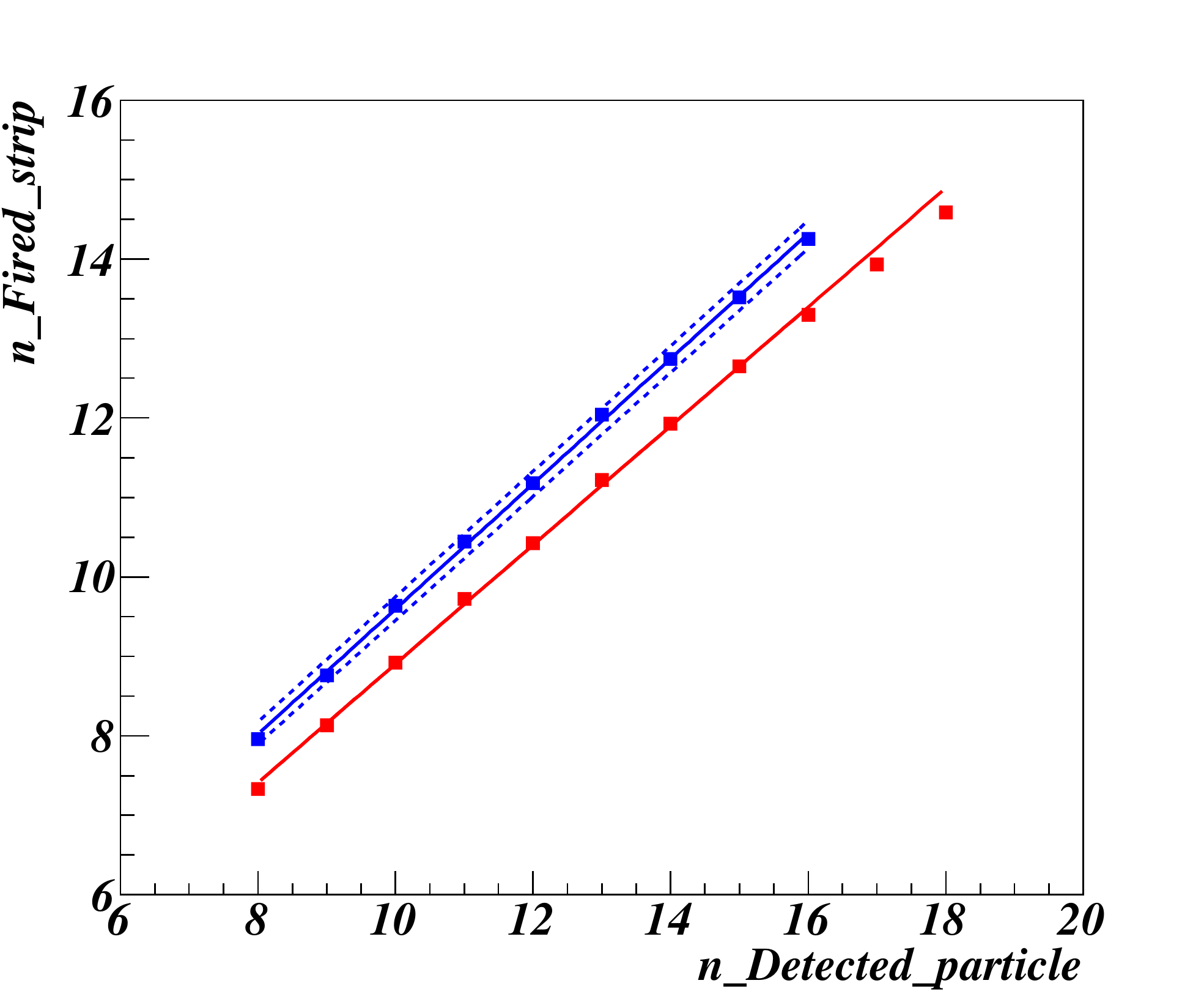}
\caption{Relation between the number of fired strips and the number of detected particles, with (blue squares) and without (red squares) induction on adjacent strips as it results from Montecarlo simulations. Blue squares refer to strip multiplicity 1.1 (that is 10$\%$ probability of signal induction on adjacent strip), the band (dashed lines) corresponds to a variation of the strip multiplicity between 1.08 and 1.12 ($\pm$ 1 s.d.).}
\label{Relation_StripVsParticle}
\end{figure}

\begin{figure}[t]
\centering
\includegraphics[width=4.0in]{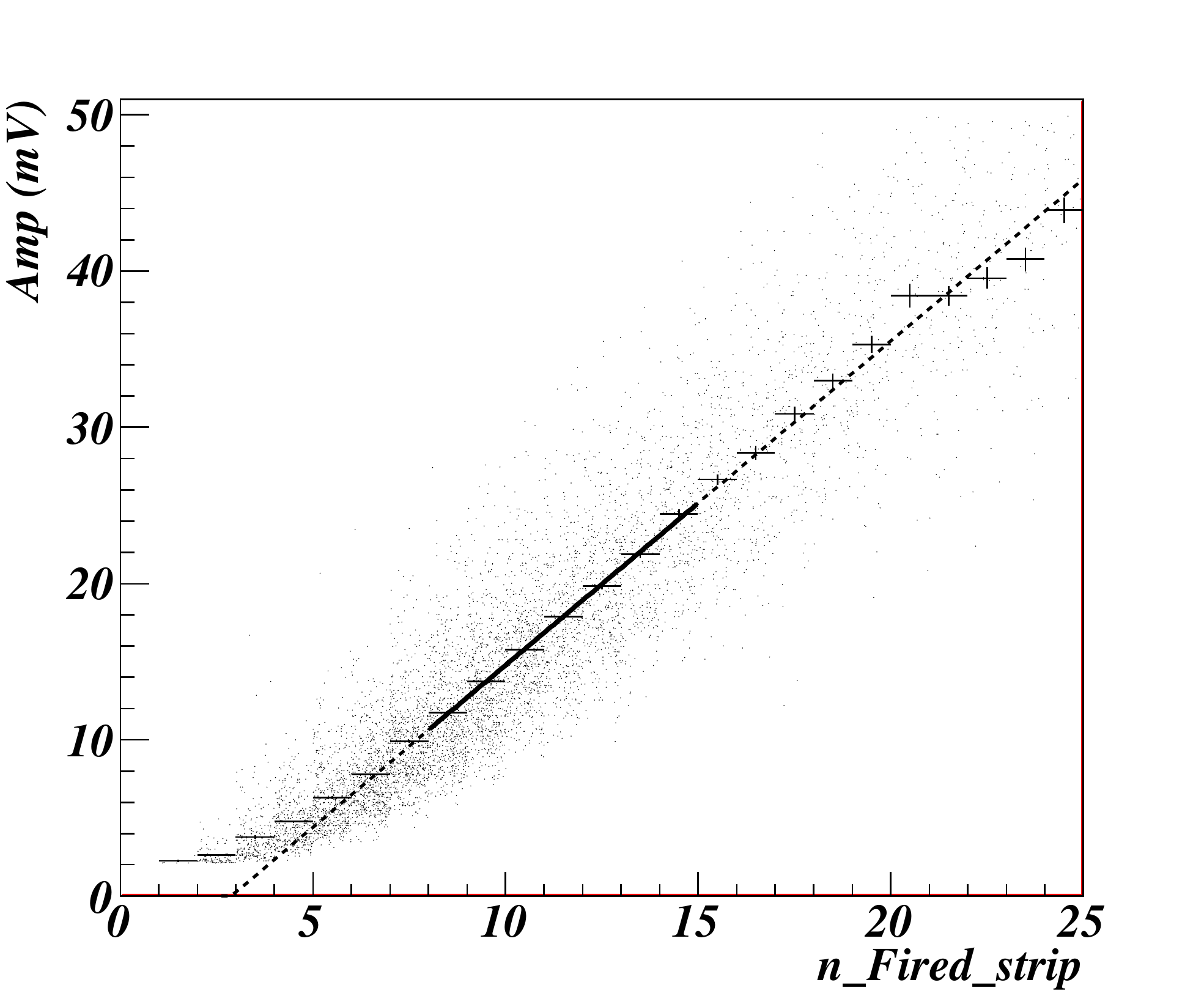}
\caption{BP amplitude versus number of fired strips. To obtain the gain a linear fit has been performed in the range 8-15 strips (heavy line). The slope of the fitting line defines the gain G(mV/strip).  }
\label{Qvolt_vs_strip}
\end{figure}

The gain calibration has been performed by using the data taken with f.s. .29 V, which is the most sensitive one and allows a direct comparison between analog and digital measurements of the same quantity, that is the number of particles hitting the RPC. Vertical showers ($\theta \leq 15^\circ$) have been selected. The adopted calibration procedure relies on the number of fired strips as a good estimate of the number of particles crossing the detector.
The good linearity between the number of fired strips and the number of detected
particles has been verified by simple simulation of the detector response.
To include the signal induction on adjacent strips (as expected in
case of particle crossing nearby the strip edge) a strip multiplicity 1.10$\pm$0.02 per
particle, as directly measured at the experiment site, has been
assumed in the simulation. The Fig.\ref{Relation_StripVsParticle} shows the dependence of the number of fired  strips on the number of detected particles in the range
of interest to the gain calibration,
the shown band corresponds to a variation of the strip multiplicity between 1.08 and 1.12.
The typical plot of the BP signal amplitude (Amp) versus the number of fired strips is
presented in Fig.\ref{Qvolt_vs_strip}: a good linearity is shown up to 15-18 fired strips,
while a clear deviation can be seen even below 7-8 fired strips.
Accordingly a linear fit was performed in the range 8 to 15 strips
that maximize the overlap region with linear response,
so defining the BP gain (G(mV/strip)) as the slope of the fitting line:
\begin{equation}
G (mV/strip)= \frac{\Delta Amp(mV)}{\Delta n_{fired-strip}}  \label{for13}; \\
\end{equation}
in case of dead strips in the BP area the fit range has been suitably re-scaled.
This fitting procedure has been applied to calibrate each BP. The number of collected events is enough to reach a statistical uncertainty in the gain determination lower than 2$\%$ That can be achieved with about 1 hour of data taking.
\begin{figure}[t]
\centering
\includegraphics[width=4.0in]{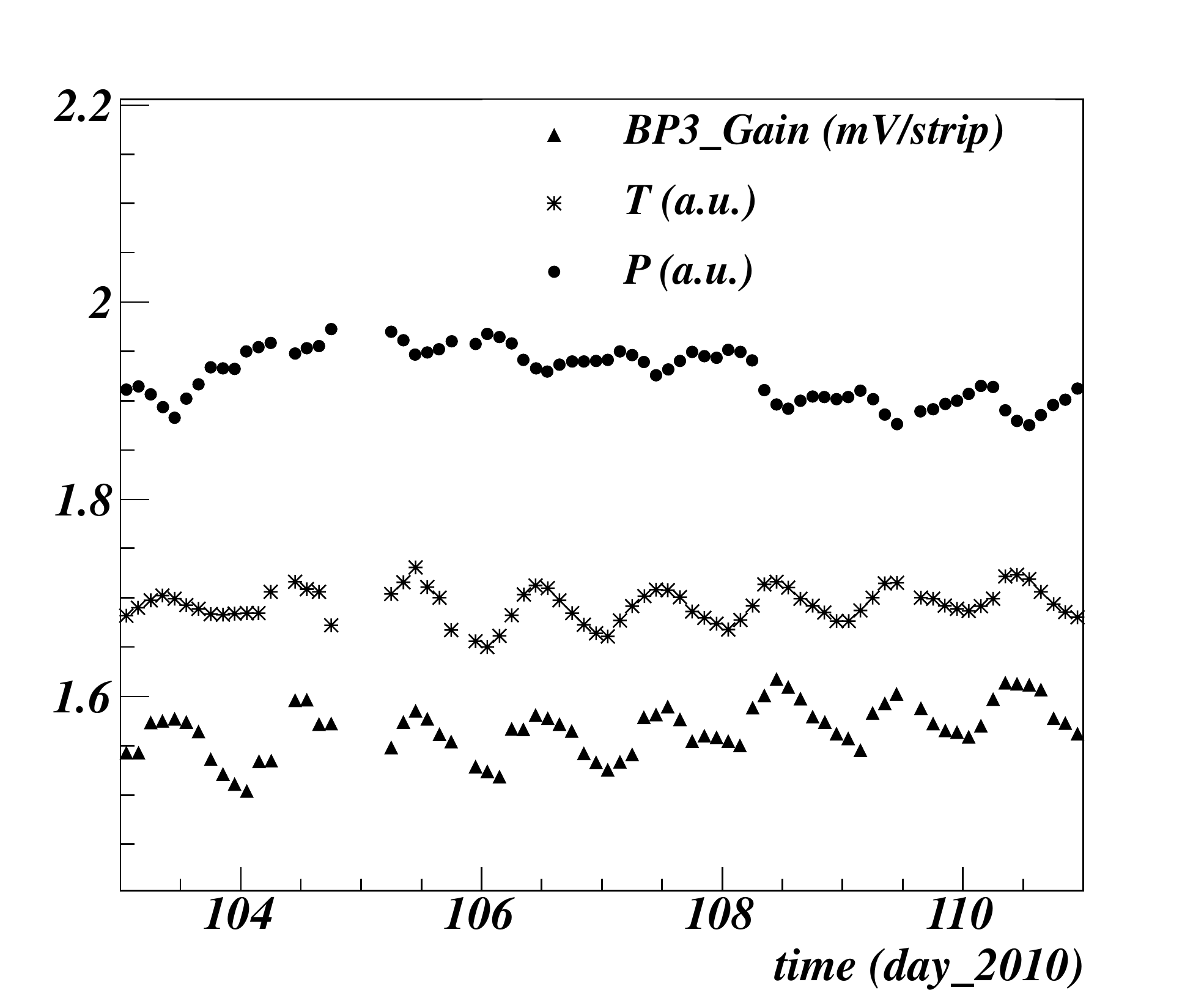}
\caption{The BP3 gain (mV/strip) monitored throughout 8 days; temperature and pressure, suitably rescaled (a.u.) to fit the gain scale, are also reported.}
\label{monBP2_P_T}
\end{figure}

Then, taking into account the relation between fired strips and detected particles, shown in Fig.\ref{Relation_StripVsParticle}, the gain G can be redefined and expressed in mV/particle:
\begin{equation}
G (mV/particle)= \frac{\Delta Amp(mV)}{\Delta n_{fired-strip}} \times \frac{\Delta n_{fired-strip}}{\Delta n_{detected-particle}}  \label{for3} \\
\end{equation}

In the following, since the relation shown in Fig.\ref{Relation_StripVsParticle} applies to all BPs, G is typically reported in mV/strip.\\
The gain distributions of BP0 and BP3 have been reported in Fig.\ref{bichambergas}.
Dividing the data sample in 1 hour sub-samples and repeating the procedure all over the sub-samples, the gain variation with time has been obtained. The dependence on  atmospheric pressure (P) and temperature (T) has been investigated. These environmental parameters are measured with an high accuracy, respectively of $\pm 0.5 \rm{mbar}$ and $\pm 0.25 \rm{K}$. In order to display the gain dependence on P and T, the mean gain has been calculated for each group of BPs having the same position along the gas line.
As an example, the behavior of the mean gain of BP3, along 8 contiguous days, is shown in Fig.\ref{monBP2_P_T}.
As expected the gain is daily modulated and correlated  to both  atmospheric pressure and temperature.

\begin{figure}[t]
\centering
\includegraphics[width=4.0in]{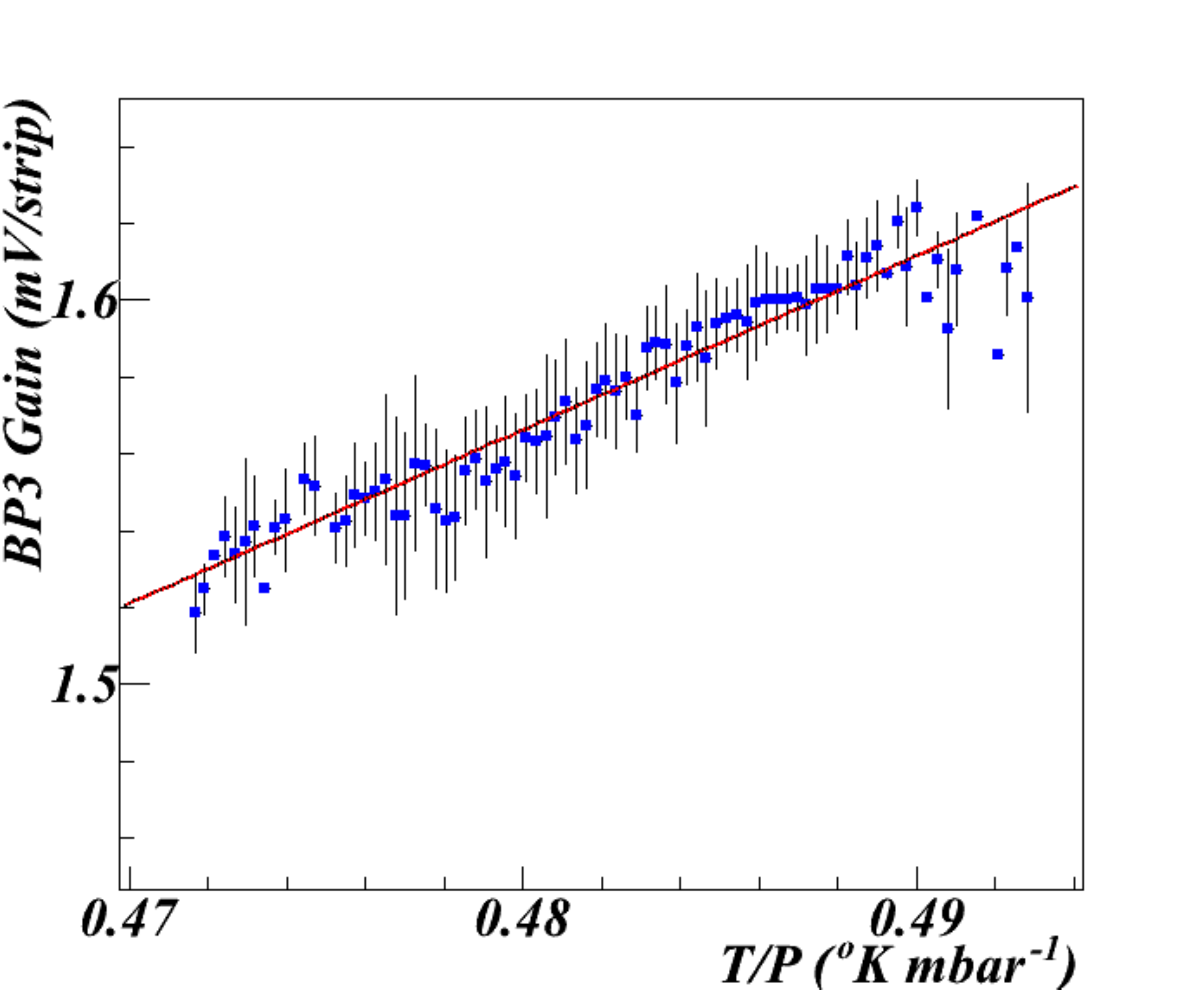}
\caption{The BP3 gain versus the T/P parameter. A linear relation fits the data quite well, it is used to correct the data for changes of the environmental parameters T and P.}
\label{Gain_vs_PT_fit}
\end{figure}

In Fig.\ref{Gain_vs_PT_fit} the best fit to the data it is shown for the relation \cite{santonico00}
\begin{equation}
G(i)=G(i)_{0} \times \frac{P_{0}}{T_{0}} \times \frac{T}{P}  \label{for1} \\
\end{equation}
where $P_{0}$ and $T_{0}$ are arbitrarily chosen and $G(i)_{0}$ is the gain of specific BP at the reference $P_{0}$ and $T_{0}$,
which has been used to correct the data for changes of the environmental parameters T and P. The same degree of correlation has been observed for BP1 and BP2, while BP0 appears  to suffer of wider variations.

\begin{figure}[t]
\centering
\includegraphics[width=5.0in]{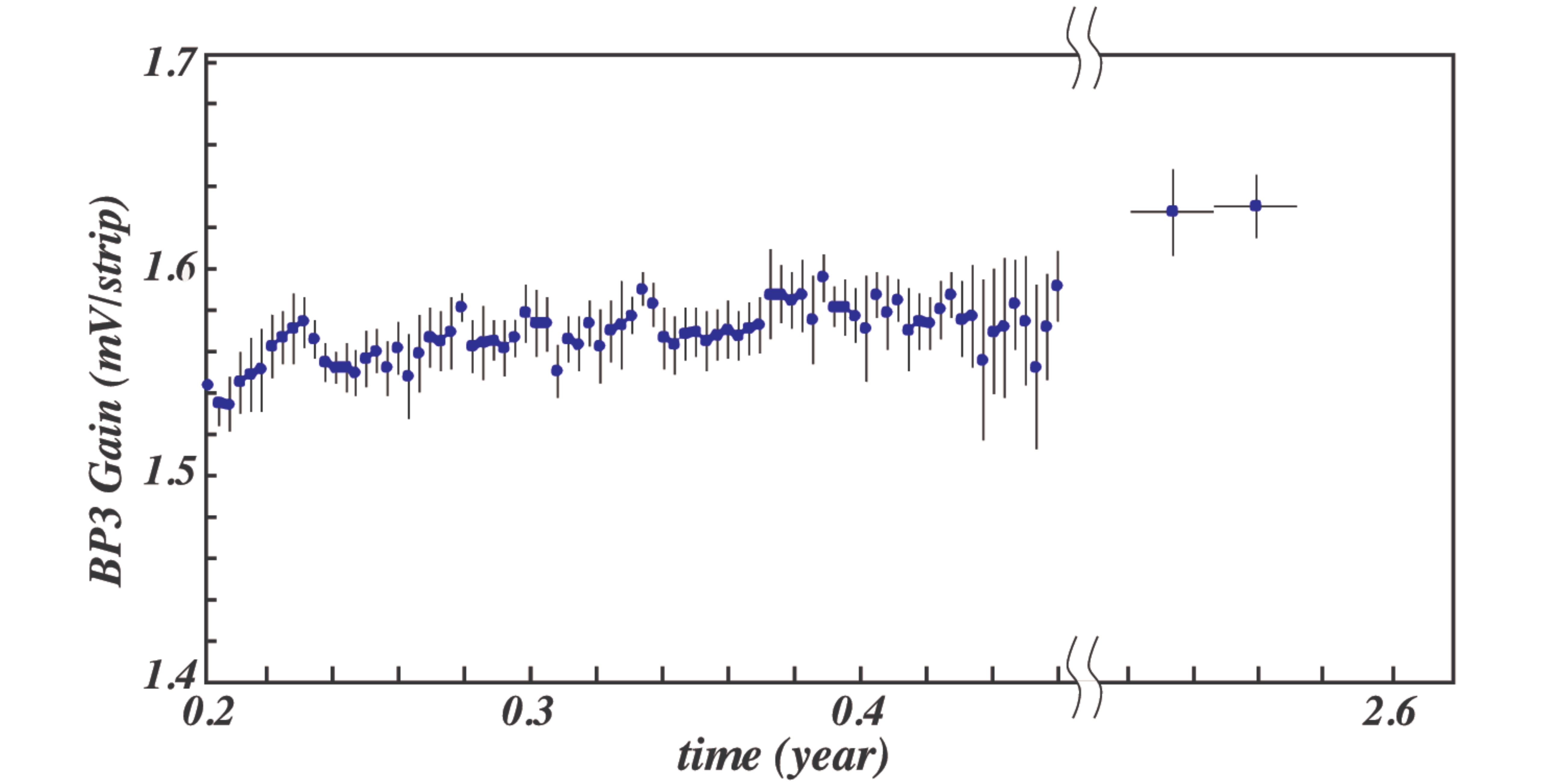}
\caption{BP3 gain behavior versus time. The horizontal scale has a break corresponding to about 2 years. The points before the break refer to about three months of data, while the last two point refer to about 3 days of data taken after 2 years. The gain variation corresponds to a derivative of the detector response of +2.5$\%$/year.}
\label{Stabilità}
\end{figure}

By correcting the gain of each BP  for the dependence on P and T, the spread of the gain distribution clearly reduces and we are left with a spread of about 2$\%$. The gain variation after correction has been used to evaluate the stability of the detector over long time. In Fig.\ref{Stabilità} the BP3 gain is reported versus time. The horizontal scale has a break corresponding to about 2 years. The points before the break refer to about three months of data, while the last two point refer to about 3 days of data taken after 2 years. The variation corresponds to a gain increase of +2.5$\%$/year; same values have been found for BP1 and BP2 while for BP0 we have  +9.5$\%$/year, therefore a mean derivative of +4.4$\%$/year due to the higher variation of BP0.

As seen, at the most sensitive scale ($\rm{R_{fs}}$=1, f.s. 0.29 V) this procedure guarantees a direct way to calibrate the analog system providing to each BP channel its own gain, on a run by run base, owing to the overlap between analog and digital information.

\subsubsection{All Scales Calibration}
At higher f.s., the procedure is quite similar: first the electronic calibration is performed, then the gain measured by the 0.29 V scale is used to extract the number of particles. The gain is taken at P$_0$ and T$_0$ which correspond to the mean values of the environmental parameters in the specific run and the correction for P and T variations is applied. This procedure relies on detector stability, which is certainly true on a single run base.
The long term variations are taken into account directly by the gain modulation as obtained by the calibration
procedure, while for very long times also the gain derivative with time, shown in Fig.\ref{Stabilità}, is considered.
All information needed to extract the number of particles ($n$) hitting the specific BP, that is electronics calibration parameters and gain of each channel, along with the correction for pressure and temperature, is managed through a Data Base.
Given the amplitude A of the measured signal (in mV) and the gain $G$ (mV/part), the number of particles hitting the specific BP is simply obtained as $ n= A/G$ with an error that, assuming A and G do combine linearly,  can be written as:
\begin{equation}
\frac{\sigma_{n}}{n}={{\frac{\sigma_{A}}{A}} +  \frac{\sigma_{G}}{G}} \label{for4} \\
\end{equation}
According to the described procedure and to the equation (\ref{for3}), many sources contribute to the error on $G$, namely:
\begin{enumerate}
  \item {the error in the fitting procedure for the mV/strip determination ($\epsilon_{As}$);}
  \item {the error coming from the strip versus particle correlation function ($\epsilon_{sp}$);}
  \item {the error on the electronic calibration ($\epsilon_{eCal}$);}
  \item {the error on the P/T correction ($\epsilon_{PTcor}$).}
\end{enumerate}
The specific errors for each source, both statistical and systematic,  are conveniently summarized in Tab.\ref{SigmaGsuG}. Therefore for the gain $G$ of the BP we have
$\sigma_{G}/{G} = (2.3\%)_{stat} + (3.8\%)_{sys}$.
\begin{table}
\begin{center}
\begin{tabular}{||c|c|c||}
\hline \hline
\small{\backslashbox{source}{error($\%$)}}  & \small{stat.} & \small{sys.} \\
\hline
\ $\epsilon_{As}$  &  {2.0}  &   {2.5} \\
\ $\epsilon_{sp}$  &  {0.9}  &  {1.3 } \\
\ $\epsilon_{eCal}$ &  {0.5} &   {1.5} \\
\ $\epsilon_{PTcor}$ &       &  {2.0} \\
\hline
\ $\epsilon_{ G }$     &  {2.3}  &  {3.8} \\
\hline \hline
\end{tabular}
\end{center}
\caption {\small{Error sources ($\%$) to the determination of $G$ }}
\label{SigmaGsuG}
\end{table}

Summing in quadrature statistical and systematic errors we get  4.4$\%$ as total error on the gain.
As far the amplitude, given the r.m.s. of the single m.i.p. signal (Sec. II) and the linearity of the RPC response (Sec. III), we have \\
\begin{equation}
{\frac{\sigma_{A}}{A}}= \frac{20\%}{\sqrt{n}} \label{for5} \\
\end{equation}
provided the measurement error on A is small, like in our case. Finally we have
\begin{equation}
{\frac{\sigma_{n}}{n}}= \frac{20\%}{\sqrt{n}} +4.4\%\label{for6} \\
\end{equation}

\begin{figure}[t]
\centering
\includegraphics[width=4.0in]{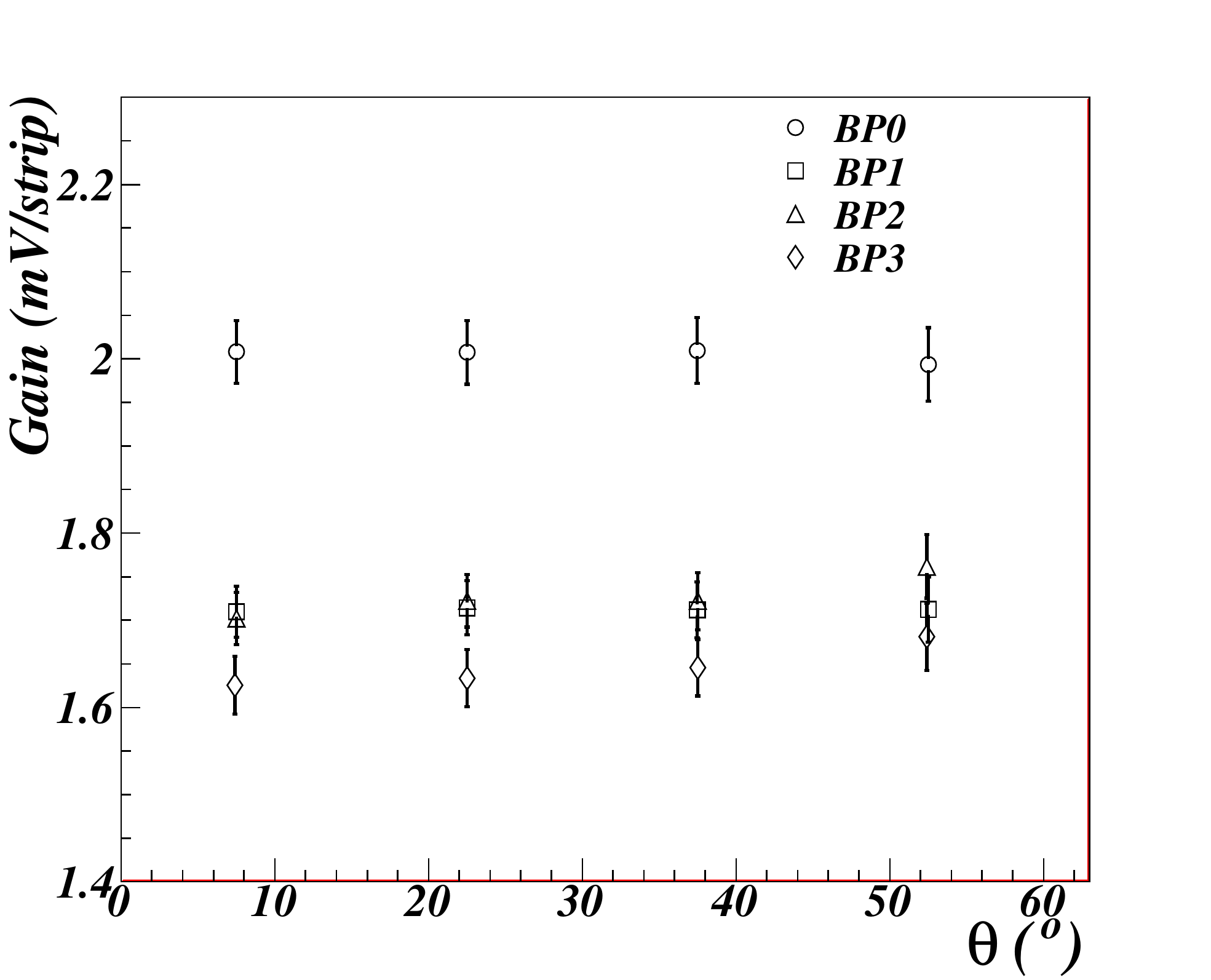}
\caption{Gain (mV/strip) versus incidence angle $\theta$ of the shower, with respect to the zenith, for the four typology of BPs.}
\label{GainVsTheta}
\end{figure}

The dependence of the gain on the zenith angle of the detected air showers has also been investigated. Data reported in Fig.\ref{GainVsTheta} show that there is no dependence on the incidence angle. Actually, given the width of the gas gap, 2 mm, and the typical size of the streamer ($\sim$3 mm), one can expect almost no dependence at least up to 50-60 degrees.

\section{Performance Evaluation.}
\begin{figure}[h]
\begin{center}
\includegraphics[width=4.0in]{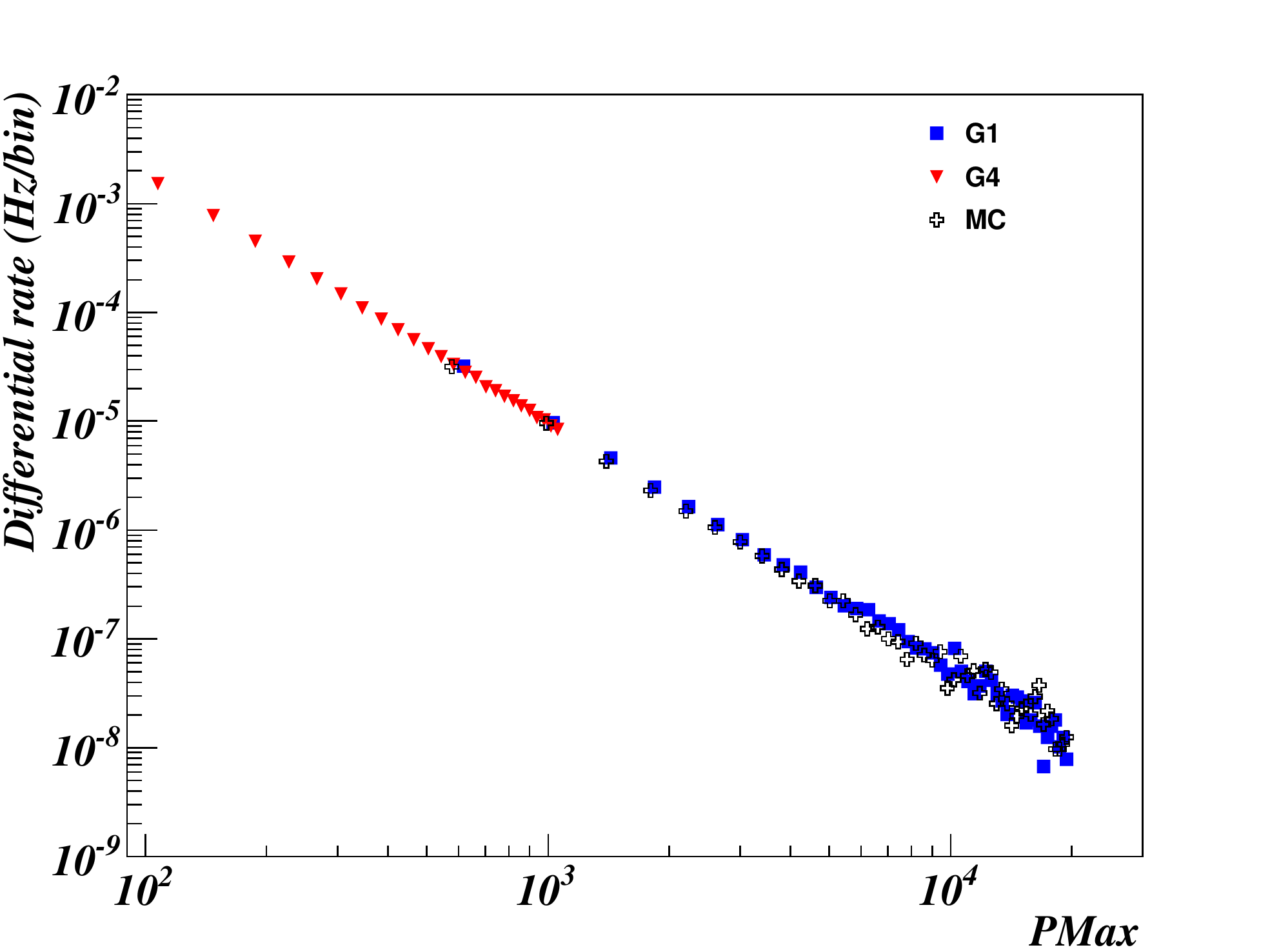}
\end{center}
\caption{Differential rate of PMax for events with core in a fiducial area of the carpet and $\theta \leq 15^\circ$,
 showing a very good matching between different scales (see text). The results from a Montecarlo simulation are shown for comparison.}
\label{fig spettro_pmax_dati}
\end{figure}

\begin{figure}[h]
\begin{center}
\includegraphics[width=4.0in]{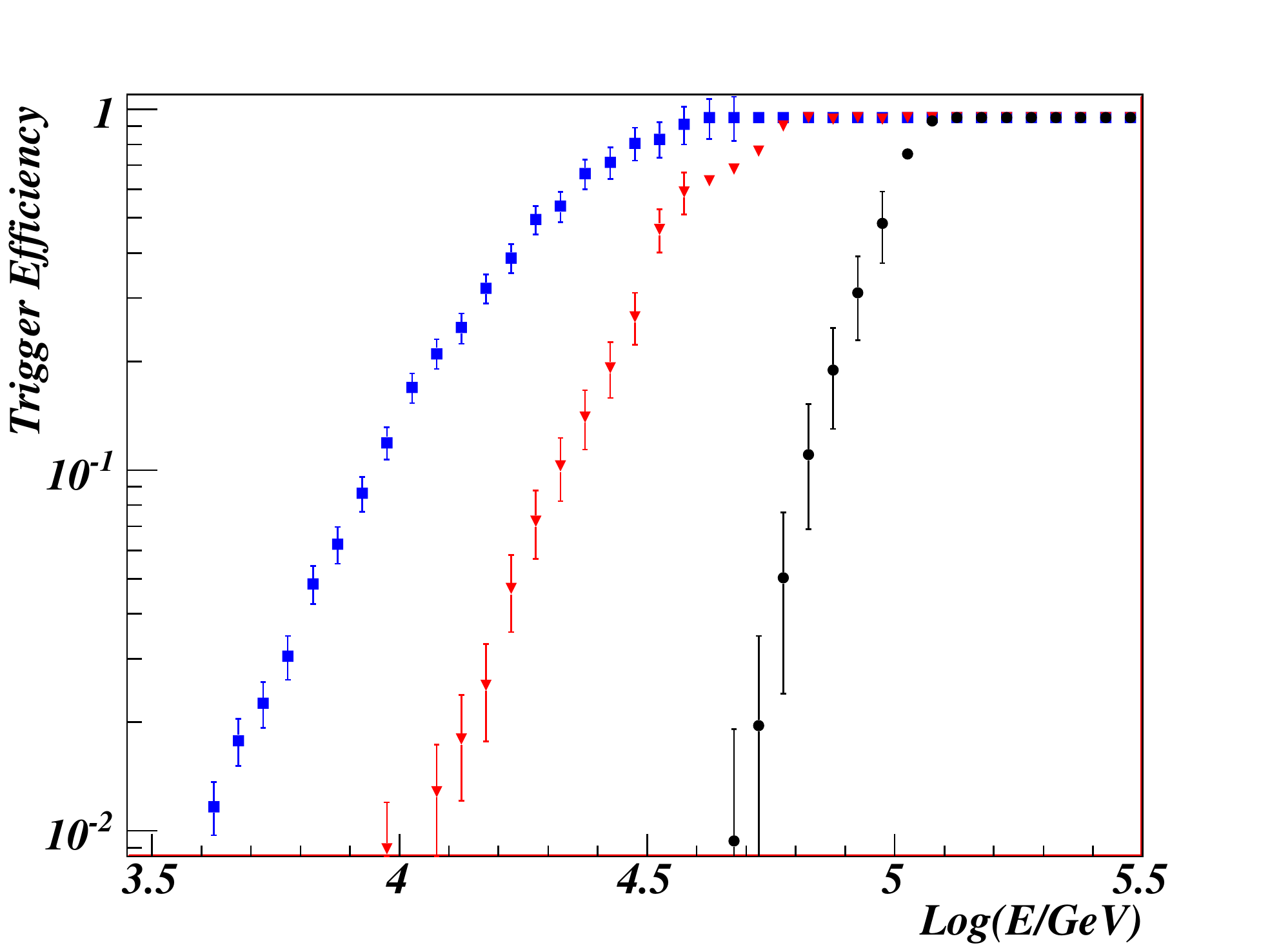}
\end{center}
\caption{The trigger efficiency of the analog system as function of the energy E of different primary cosmic rays (protons (blue squares), He (red triangles), Fe(black dots)).}
\label{TriggerEfficiency}
\end{figure}

\begin{figure}[h]
\begin{center}
\includegraphics[width=4.0in]{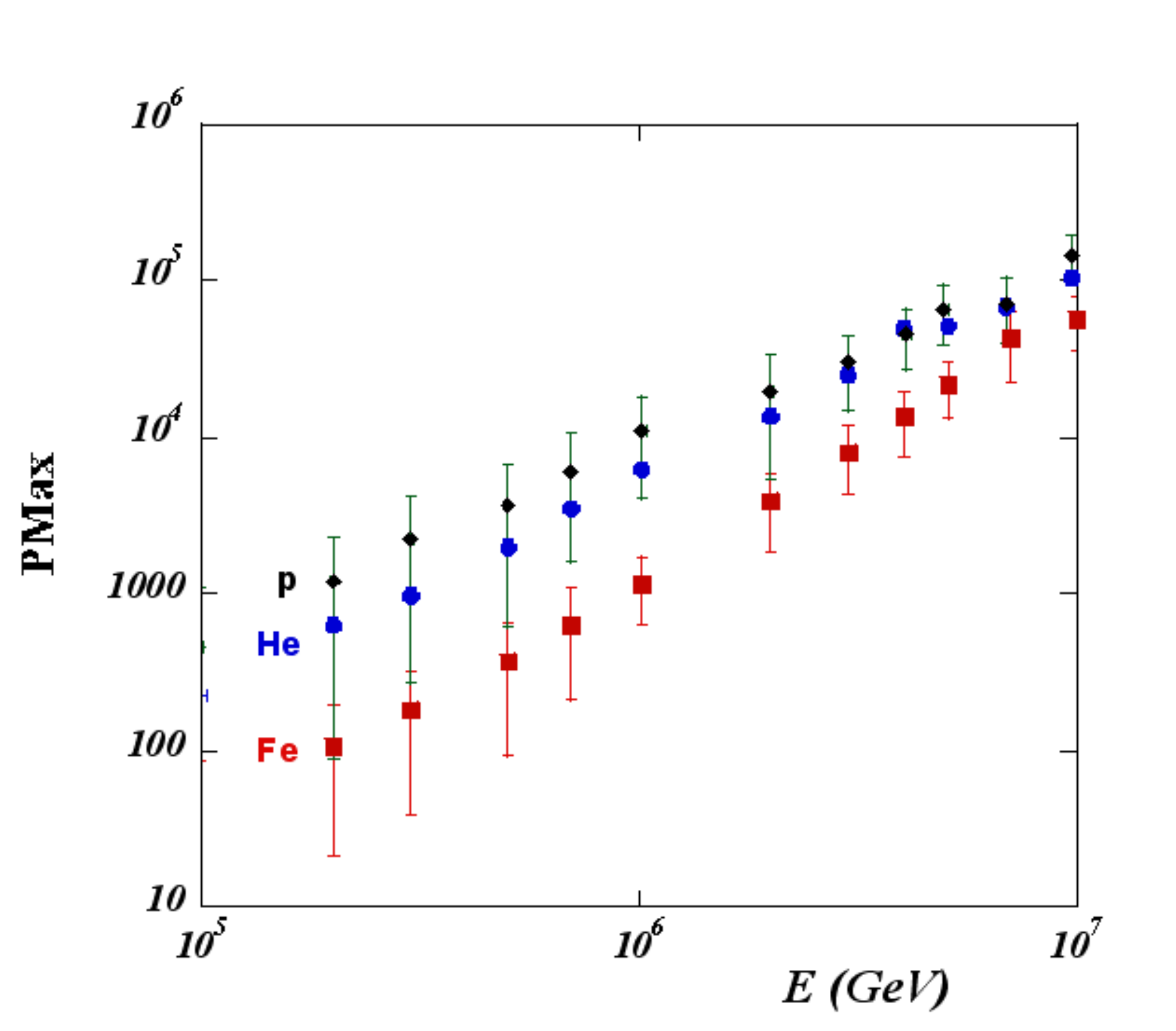}
\end{center}
\caption{Number of particles falling in the BP at the shower core position (PMax) versus energy, for different primaries, namely protons, He and Fe. The r.m.s. (error bar) of PMax is reported only for protons and Fe in order to avoid heavy overlap.}
\label{PMaxvsEnergy}
\end{figure}

In order to estimate the goodness of the calibration procedure the data concerning each BP have been analyzed at the different steps of the calibration procedure, in terms of ADC count,  amplitude and number of particle. The distribution of each of these quantities, on a single BP, follows a power law as a consequence of the shape of the primary cosmic ray spectrum. Therefore, by fitting the distribution, the power index of the mentioned variables has been  measured for each BP. If the calibration procedure operates correctly,
then the spread of the index distribution reduces going from the raw ADC count to the reconstructed number of particles.
This check has been done for both f.s. 0.29 V and f.s 16.3 V, the results are summarized in Tab.\ref{TAB_Slope}.
Looking at values of the table a few considerations can be done, namely:
a) the spread of the slope distributions in case of the ADC count are almost independent of f.s. (ADC count row);
b) the electronic calibration (Ampl row) seems to perform better at higher f.s. but this is due just to the calibration procedure, in fact more points are used at higher f.s.;
c) in case of strips (particles) the slopes and their r.m.s. values are quite independent of the scale.
Finally, according to the last row, 4.3$\%$ is a good estimate of the homogeneity among BPs after calibration.
\begin{table}
\begin{center}
\begin{tabular}{||c|c||c||}
\hline \hline
\small{$<>\pm rms$}  & \small{f.s. .29 V} & \small{f.s. 16.2 V} \\
\small{($rms/<>$)}     &                &  \\
\hline
\small{ADCcou} & \small{$2.39\pm 0.19$ $(8.2\%)$} & \small{$1.90\pm 0.18$ $(9.4\%)$} \\
\small{Ampl}   & \small{$2.55\pm 0.18$ $(7.1\%)$} & \small{$2.56\pm 0.13$ $(5.1\%)$} \\
\small{Part}  & \small{$2.51\pm 0.11$ $(4.4\%)$} & \small{$2.51\pm 0.11$ $(4.2\%)$} \\
\hline \hline
\end{tabular}
\end{center}
\caption {\small{Index of the power law distribution of ADC count (ADCcou), Amplitude (Ampl) and number of particles (Part) in a BP: The mean value and the r.m.s. of the index distribution (see text for details), for all BPs, in case of f.s. 0.29 V and f.s. 16.2 V, are reported.}}\label{TAB_Slope}
\end{table}
\\
In order to check any systematic effects related to the use of different scales, which means the correct matching between different energy ranges, events with core in a fiducial area of the central carpet (the $6 \times 9$ central Clusters, corresponding to 2380 m$^2$) and zenith angle $\theta \leq 15^\circ$ have been selected and the distribution of the number of particles on the BP at the shower core position (PMax), or the BP with the highest signal, has been studied.
PMax is correlated to both energy and mass of the primary particle. The differential rate of Pmax is reported in Fig.\ref{fig spettro_pmax_dati} for the two scales of operation G4 and G1.
It can be seen a perfect matching between contiguous ranges and absence of any systematic effect. We notice that Pmax spans over two and half decades, while the event frequency runs over five decades.\\
An absolute comparison of the experimental distribution in Fig.\ref{fig spettro_pmax_dati} has been done with Montecarlo expectations.
For this aim, showers have been generated by the Corsika \cite{Knapp93} code with QGSJETII as hadronic interaction model. Energies have been generated according to the spectra in \cite{Horand03} for proton, and up to Fe, primaries.  The induced showers have been sampled at the YangBaJing elevation. The shower cores were distributed in an area wider than the fiducial area of the central carpet ($64 \times 64$ m$^2$) and  almost vertical showers have been considered ($\theta \leq 30^\circ$). Detector and trigger have been fully simulated.
The same cuts used for the real event selection were applied to the Montecarlo sample. The Montecarlo expectation for  PMax is reported in Fig.\ref{fig spettro_pmax_dati} for an absolute comparison with data, showing a very good agreement.
The same events have been used to estimate the energy threshold of the analog trigger with respect to different primaries.
The trigger efficiency versus energy, for different primaries, namely protons, He and Fe is reported in Fig.\ref{TriggerEfficiency}, it shows that above 100 TeV the analog system is fully efficient for all primary particles.
The relation of PMax to the energy of the primary has been obtained by the above mentioned Montecarlo simulations at fixed energies, the results are reported in Fig.\ref{PMaxvsEnergy} for protons, He and Fe. Considering the RPC intrinsic linearity, which has been demonstrated at least up to particle density of 2x10$^4$ m$^{-2}$, and the electronics features, this implies we can fully explore the CR spectrum up to $\sim$ 10 PeV. The related statistics of detectable events, per year of operation and vertical events, is shown in Tab.\ref{EvtStat} for different primaries .

\begin{table}
\begin{center}
\begin{tabular}{||c|c|c|c||}
\hline \hline
\small{\backslashbox{E$_{th}$ (TeV)}{Primary}}  & \small{p} & \small{He} & \small{Fe} \\
\hline
\small{100}   & \small{3.08e+05}    & \small{2.89e+05}   & \small{1.35e+05} \\
\small{500}   & \small{1.81e+04 }   & \small{2.02e+04}   & \small{1.04e+04} \\
\small{1000}  & \small{4.82e+03}    & \small{6.07e+03}   & \small{3.45e+03} \\
\small{2000}  & \small{1.07e+03}    & \small{1.68e+03}   & \small{1.14e+03} \\
\small{4000}  & \small{1.70e+02}    & \small{3.91e+02}   & \small{3.75e+02} \\
\small{10000} & \small{8.5}         & \small{3.37e+01}   & \small{ 8.2e+01} \\
\hline \hline
\end{tabular}
\end{center}
\caption {\small{Number of vertical ($\theta < 15\deg$) events per year of operation in a fiducial area of $\sim$ 2400 m$^2$ for $\rm{E>E_{th}}$} and different primaries.}
\label{EvtStat}
\end{table}

\section {Summary}
In this paper the analog detector of the ARGO-YBJ experiment and its performance have been presented.
The features of the RPC signals are discussed within the framework of the DAQ of the experiment.
Results on the intrinsic linearity of the RPCs, as measured at the BTF of the INFN National Laboratory in Frascati (Italy), are reported. These results confirm a linear behavior up to  particle density of $2\times10^4$ $m^{-2}$, at least.
The analog detector is able to operate at different energy ranges, owing to the flexibility of the readout electronics. At the most sensitive scale, the digital readout and the analog readout perfectly overlap so providing a powerful instrument of calibration. The calibration procedure is described in all details; it is self consistent because relies on two independent measurements of the same quantity, being the RPC signals independently readout by the strips and by the analog pickup electrode (BP). No additional information is needed.
The error on the measurement of the number of particles (n) crossing the BP is essentially determined by the intrinsic fluctuation of the RPC signal, which is about 20$\%$ for a single streamer at the operational conditions of the experiment, while decreases as $1/\sqrt(\rm{n})$. The contribution to the error of the electronic calibration  is less than 0.5$\%$, while the electronics itself shows a quite good stability, or a seasonal variation of the response which is less than 1$\%$.
The mean signal of the single particle, the gain, is determined with an uncertainty of about $\rm{(2.3\%)_{stat} + (3.5\%)_{sys}}$ on the single run base, or a few hours; its mean time derivative is  $\rm{+ (4.3\%)/year}$  which also represents the detector stability.
More than 95$\%$ of the 3120 BP in the central carpet of ARGO-YBJ have been calibrated with the described procedure.
The study of the trigger efficiency has confirmed the analog detector to be fully efficient above 100 TeV for all kind of primary cosmic rays.
The physics performance of the detector has been evaluated by studying the number of particles hitting the BP at the shower core position (PMax). The spectrum of this variable, as obtained by data analysis of showers with core in a fiducial area of the ARGO-YBJ carpet, has shown no systematic effects related to the use of different scales, that is a perfect matching in the overlapping regions of this distribution.
An absolute comparison of the distribution with the Montecarlo expectation shows a very good agreement within the experimental uncertainties.
The correlation of PMax with the energy, for different primaries, has been investigated, thus confirming that ARGO-YBJ can efficiently operate up to PeV energies beyond the knee of the primary cosmic ray spectrum.

\section {Conclusion}

As in the case with all air shower detectors, ARGO-YBJ samples the cascades induced deep in atmosphere by primary cosmic rays. Information about primary composition and energy spectra is inferred from measurements of the properties of the secondary air showers. The analog readout of the ARGO-YBJ RPC detector, put in operation on December 2009, allows for a very detailed and precise measurement of the number of charged particles around the core of air showers induced by $>$ 100 TeV primaries. This imaging is achieved by means of an array of 3120 pixels ( called BigPad) of 1.7 m$^2$ each, densely distributed over a 93$\%$ active area of about 5770 m$^2$ . The most important features of this device can be summarized as :
\begin{itemize}
\item {each pixel is instrumented to count the number n of charged particles with a resolution 20$\%/\sqrt(n)$ + 4.4$\%$ where the first term is related to the fluctuation of the streamer formation in the gas gap of the RPC , while the second term accounts for the uncertainty in the absolute calibration functions and control of the environmental parameters. As far the long-term stability of the detector, the m.i.p. signal shows a mean variation of of +4.3$\%$/year;}
\item {the dynamic range of each pixel extends as far as particle densities of 2$\times 10^4 /m^2$ and more, allowing to measure the number of charged particles around the core of air shower from PeV primaries with an accuracy $\leq 5\%$;}
\item {the performance of this device does not depends on the air shower incidence angle at least up to 50 degrees.}
\end{itemize}
In consequence, since the position of the detector at high altitude makes possible to sample ground particles near the shower maximum, thus reducing significantly the effects of fluctuations, one can reasonably predict accurate measurements around the knee of the primary cosmic ray spectrum with unprecedented resolution.
Preliminary results presented at ICRC 2013 \cite{iacovacciICRC13} as well as the study of the PMax distribution discussed in this paper confirm the excellent performance of the detector and of its analog readout.

\section*{Acknowledgment}
This work is supported in China by NSFC, the Chinese Ministry of Science and Technology, the Chinese Academy of Sciences, the Key Laboratory of Particle Astrophysics, CAS, and in Italy by the Istituto Nazionale di Fisica Nucleare (INFN) and, partially, by Ministero per gli Affari Esteri (MAE), Government of Italy.
We also acknowledge the essential support of W.Y. Chen, G. Yang, X. F. Yuan, C.Y. Zhao, R. Assiro, B. Biondo, S. Bricola,
F. Budano, A. Corvaglia, B. D'Aquino, R. Esposito, A. Innocente, A. Mangano, E. Pastori, C. Pinto, E. Reali, F.
Taurino, and A. Zerbini in the installation, debugging, and maintenance of the detector.
We thank the group of the DAFNE Beam-Test Facility, especially G. Mazzitelli and P. Valente, for their valuable support during and after the test beam.

\end{document}